%% file: ms.tex
\documentclass{aa}

\usepackage{color}
\usepackage{hyperref}
\usepackage[varg]{txfonts}
\usepackage{calrsfs}
\usepackage[normalem]{ulem}
\DeclareMathAlphabet{\pazocal}{OMS}{zplm}{m}{n}

\hypersetup{
    colorlinks = true,
    citecolor = blue,
    linkcolor = blue
}

\newcommand{\ra}[1]{#1}
\newcommand{\ratwo}[1]{#1}

\begin{document} 

\title{Convolutional neural networks on the HEALPix sphere: a pixel-based algorithm and its application to CMB data analysis}
\author{%
N. Krachmalnicoff\inst{1,2}\thanks{e-mail: \href{mailto:nkrach@sissa.it}{nkrach@sissa.it}}%
\and%
M. Tomasi\inst{3,4}}
\institute{%
SISSA, Via Bonomea 265, 34136, Trieste, Italy%
\and%
IFPU, Institute for Fundamental Physics of the Universe, Via Beirut 2, 34014, Trieste, Italy
\and%
Dipartimento di Fisica ``Aldo Pontremoli'', Università degli Studi di Milano, Via Celoria 16, 20133, Milano, Italy
\and%
Istituto Nazionale di Fisica Nucleare (INFN), sezione di Milano, Via Celoria 16, 20133, Milano, Italy}
\date{}

\abstract{We describe a novel method for the application of convolutional neural networks (CNNs) to fields defined on the sphere, using the Hierarchical Equal Area Latitude Pixelization scheme (HEALPix). Specifically, we have developed a pixel-based approach to implement convolutional and pooling layers on the spherical surface, similarly to what is commonly done for CNNs applied to Euclidean space. \ra{The main advantage of our algorithm is to be} fully integrable with existing, highly optimized libraries for NNs (e.g., PyTorch, TensorFlow, etc.). \\
We present two applications of our method: (i) recognition of handwritten digits projected on the sphere; (ii) estimation of cosmological parameter from simulated maps of the Cosmic Microwave Background (CMB). \ra{The latter represents the main target of this exploratory work, whose goal is to show the applicability of our CNN to CMB parameter estimation.} We have built a simple NN architecture, consisting of four convolutional and pooling layers, and we have used it for all the applications explored herein.\\
Concerning the recognition of handwritten digits, our CNN reaches an accuracy of $\sim95\%$, comparable with other existing spherical CNNs, \ra{and this is true regardless of the position and orientation of the image on the sphere.} For CMB-related applications, we tested the CNN on the estimation of a mock \ra{cosmological} parameter, defining the angular scale at which the power spectrum of a Gaussian field projected on the sphere peaks. We estimated the value of this parameter directly from simulated maps, in several cases: temperature and polarization maps, presence of white noise, and partially covered maps. \ra{For temperature maps,} the NN performances are comparable with those from standard spectrum-based Bayesian methods. \ra{For polarization, CNNs perform about a factor four worse than standard algorithms. Nonetheless, our results demonstrate, for the first time, that CNNs are able to extract information from polarization fields, both in full-sky and masked maps, and to distinguish between $E$ and $B$-modes in pixel space.}\\
Lastly, we have applied our CNN to the estimation of the Thomson scattering optical depth at reionization ($\tau$) from simulated CMB maps. Even without any specific optimization of the NN architecture, we reach an accuracy comparable with standard Bayesian methods. This work represents a first step towards the exploitation of NNs in CMB parameter estimation and demonstrates the feasibility of our approach.}
 
\titlerunning{Convolutional Neural Networks on the HEALPix sphere}
\maketitle
 
 \input{intro}
\input{section2}
\input{section3}

\input{section4}
\input{section5}
\input{section6}
\input{conclusions}

\begin{acknowledgements}
This research was supported by the he ASI-COSMOS Network (\url{http://cosmosnet.it}).  We acknowledge support from the RADIOFOREGROUNDS project, funded by the European Commission's H2020 Research Infrastructures under the Grant Agreement 687312. The authors thank Prof. Carlo Baccigalupi for useful discussion and early paper review. We acknowledge the use of NERSC for the simulations done in this work.
\end{acknowledgements}

\bibliographystyle{aa}
\bibliography{ms}

\appendix
\input{appendix_striped_maps}

\end{document}

%% file: intro.tex
\section{Introduction}

The field of astrophysics and cosmology is experiencing a rapidly increasing interest in the use of machine learning (ML) algorithms to analyze and efficiently extract information from data coming from experiments. This is happening because of two main factors: (i) the more and more extensive size of astronomical datasets and the ever-increasing requirements in instrumental sensitivities encourage the development of new algorithms to do the job; and (ii) ongoing research in ML continuously produces new concepts, approaches, algorithms, to classify and regress data. ML methods can demonstrate outstanding performance, and the astrophysical community is promptly taking the challenge to adapt them to the problems typically tackled in the field.

One of the most fruitful fields of ML is the domain of artificial neural networks (or neural networks for short, NNs). Roughly speaking, NNs are non-linear mathematical operators that can approximate functions that map inputs onto outputs through a process called \emph{training}. NNs have already been applied in many fields of science, such as biology \citep{Cartwright2008, Lodhi2012, Angermueller2016}, economics \citep{Lam2004, Kim2007, Yan2017, Tamura2018}, medicine \citep{Virmani2014, Kasabov2014, Cha2016, Liu2017, Cascianelli2017}, and of course astrophysics \citep{Collister04, Dieleman15, Baccigalupi2000, Graff14, Auld07}.

In this paper, we present a novel algorithm that extends commonly-used NN architectures to signals projected on the sphere. Spherical domains are of great interest for the astrophysical and cosmological community, which often observe vast portions of the celestial sphere whose curvature is not always negligible. The latter statement applies to several fields, for example: (i) the study of the Cosmic Microwave Background (CMB), the relic thermal radiation coming from the early Universe, which carries crucial information concerning very high energy physical processes occurred just after the Big Bang, and (ii) the analysis of the Universe Large Scale Structure, that is, the mapping of the location and properties of galaxies over large portions of the sky, which yields fundamental clues about the dark cosmological components.

Given our background in the analysis of data coming from experiments observing the CMB radiation, we propose applications of our algorithm to the problem of estimating cosmological parameters from CMB observations. \ra{While we provide only a proof-of-concept of such an application, this work demonstrates for the first time the feasibility of using deep CNNs for CMB parameter estimation on the very large angular scales, in both temperature and polarization.} An implementation of the proposed algorithm, developed in Python, is available at \url{https://github.com/ai4cmb/NNhealpix}.

The paper is structured as follows: in Section~\ref{sec:NN}, we introduce the basic concepts underlying the way standard and convolutional NNs work; in Section~\ref{sec:HealpixNN}, we explain the problem of the application of convolutional NNs on signals projected on the sphere and present our algorithm. Section~\ref{sec:env} briefly describes the working environment used to develop and run our algorithm, as well as the NN architecture we have used in our analysis. Sections~\ref{sec:MNIST} and~\ref{sec:CMBstuff} deal with possible applications. In Section~\ref{sec:MNIST}, we show the performance of our algorithm on a classification problem (recognition of hand-written digits from the MNIST dataset) on a spherical domain. Section~\ref{sec:CMBstuff} describes the application of our algorithm to CMB-related examples, namely the estimation of cosmological parameters applying CNNs to simulated CMB maps. Lastly, in Section~\ref{sec:conclusions} we discuss our main results and draw our conclusions.

%% file: section2.tex
\section{Basic concepts of neural networks}
\label{sec:NN}

In this section, we provide a brief introduction to the basic concepts of NNs. We stick to the two most common types of NNs, which are fully-connected networks and convolutional neural networks (CNNs).

A NN is a computational tool able to approximate a non-linear function $f$ that maps some inputs $\mathbf{x}\in \mathbb{R}^n$ into outputs $\mathbf{y}\in \mathbb{R}^m$: $\mathbf{y} = f(\mathbf{x})$. The advantage of NNs is the lack of requirements on any \textit{a priori} information for the function $f$, as the network can learn it from a (sufficiently large) set of labeled data, that is, a \emph{training set} of inputs for which the corresponding outputs are known. Given this requirement, NNs fall into the category of supervised machine learning algorithms. NNs can approximate even extremely complicated mappings by recursively applying a non-linear function (called \emph{activation function}) to linear combinations of the input elements. 

\subsection{Fully connected NN}
\label{sec:NNfc}

Several network architectures have been developed in the last years; a particularly simple type builds on so-called \emph{fully connected layers}. In this kind of NN, the architecture tries to approximate some function $f: \mathbb{R}^n \rightarrow \mathbb{R}^m$ via a sequence of multiple groups of \emph{artificial neurons}; each group is called \emph{layer}. Neurons of consecutive layers are linked to each other by \emph{connections}. Each connection between neuron $i$ in layer $K$ and neuron $j$ in layer $K+1$ is quantified by a scalar weight $w^{K(K+1)}_{ij}$, with $w^{K(K+1)}_{ij} = 0$ if the two neurons are not connected directly. The first layer of a NN, called the \emph{input layer}, contains $n$ neurons and is fed with the input.  The last layer, the \emph{output layer}, produces the result of the computation and contains $m$ neurons. All the other layers, sandwiched between the first and last, are called \emph{hidden layers}. NNs are called \emph{shallow} if they have few hidden layers, or \emph{deep} if they contain several layers (up to a thousand and more \citep{He2016}).\par

The network processes the input and produces the output via the following procedure:
\begin{enumerate}
	\item the $n$ elements of the input vector $\mathbf{x} \in \mathbb{R}^n$ are associated to each neuron in the input layer ($K=0$);
	\item for every neuron $j$ in the $(K+1)-th$ hidden layer, the following value is computed\footnote{The formula in Eq.~\protect\eqref{eq:nn} is called the weighted-average formula, and it is the most common. Depending on the domain and the purpose of the NN, other formulae can be used.}, considering all the neurons of the $K$-th layer pointing to $j$:
	\begin{equation}
	\label{eq:nn}
	\mathrm{net}^{(K+1)}_j= g_j\left(\sum_{i} \mathrm{net}^K_i \cdot w^{K(K+1)}_{ij}\right),
	\end{equation}
	where $g_j$ is the non-linear activation function for neuron $j$, $\mathrm{net}^K_i$ is the value associated with neuron $i$ of the $K$-th layer, and the sum is done over all the neurons $i$ that are connected to $j$.
	\item The output $\mathbf{\widetilde{y}} \in \mathbb{R}^m$ is the vector associated with the $m$ neurons of the output layer, obtained after the computation of the previous step in all the hidden layers.
\end{enumerate}
\par

In order for the network to be able to approximate the function $f$, the connection weights must be optimized. This is done through the network \emph{training}: starting from initialized values of the weights $w^{K(K+1)}_{ij}$, the network is fed with the elements $\mathbf{x^i}$ of the training set and the output values $\mathbf{\widetilde{y}^{\,i}}$ are obtained. A \emph{cost function} $\pazocal{J}$ is computed:
\begin{equation}
\pazocal{J}=\frac{1}{N}\sum_{i=0}^{N}\mathcal{L}(\mathbf{y^i},\mathbf{\widetilde{y}^{\,i}}),
\end{equation}
where, $\mathcal{L}$ is a \emph{loss function} that measures the distance between the true known output value $\mathbf{y^i}$  and the one computed by the network $\mathbf{\widetilde{y}^{\,i}}$, for the $i$-th element of the training set; the sum is done over $N$ elements of the training set. The minimum of the cost function is found in the following way: (i) the network runs over $N$ elements of the training set; (ii) the cost function is computed; (iii) the values of the weights are updated moving towards the minimum of the loss function, usually with a gradient descent algorithm that requires the computation of the derivative of $\pazocal{J}$ with respect to the weights; (iv) the full computation is iterated until the value of $\pazocal{J}$ converges to its minimum. The application of steps (i)--(iii) on the entire training set is called \emph{epoch}. Depending on the approach, $N$ can be equal to one, to the total number of elements in the training set, or to a sub-fraction of it (randomly changing the training set elements included in this fraction at every iteration). In the latter case, the training set is divided in so-called \emph{mini-batches}.\par

The NN parameters (e.g., the number of layers and neurons, the loss function, the dimension of mini-batches, the learning rate for the gradient descent algorithm, etc.) are optimized by evaluating the performance of the network over the \emph{validation set}  for which  the outputs $\mathbf{y}$ are also known. The final assessment of the ability of the NN to approximate the mapping $f$ is done over the \emph{test set}. Depending on the characteristics of the output and the function $f$, NNs can be used, for example, to solve classification problems or to perform parameter regression.

\subsection{Convolutional NNs}
\label{sec:CNN}

A peculiar type of NN is the so-called \emph{convolutional neural network} \citep{alexnet}, which is the main topic of this paper. In CNNs, \emph{convolutional} layers take the place of some or all of the the fully connected hidden layers. A convolutional layer convolves its input with one or more kernels (filters), and the coefficients of these kernels are the weights that are optimized during training.

\begin{figure}
	\centering
	\includegraphics{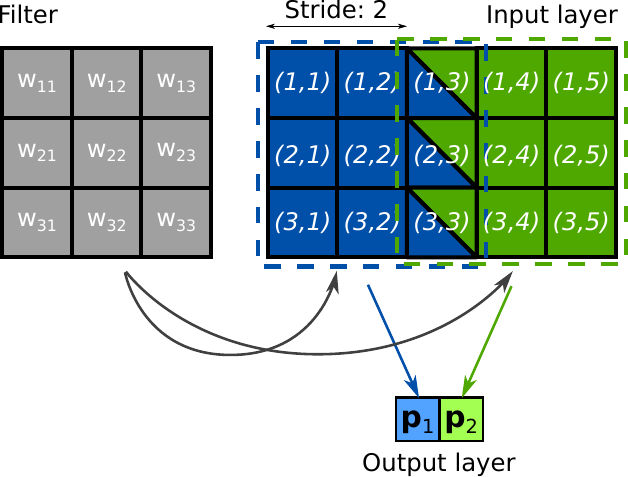}
	\caption{\label{fig:euclideanFilterOutline} Application of CNN convolutional filter to 2D image. The input image (shown as a colored matrix) contains $3 \times 5$ pixels, while the convolutional filter (in gray) is a $3 \times 3$ matrix. The filter moves across the input image and each time a dot product between the filter and the sub $3\times3$ portion of the image is performed. The result of the product is stored in the output. In this example the filter is not applied to all the possible $3\times3$ partitions of the input image but jumps one pixel at every move (the way the filter moves across the image is controlled by the \emph{stride} parameter, which in this case is equal to 2 as the filter is translated by two pixels at every move). The filter elements $w_{ij}$ represent the weights of the network which are optimized during training, and do not change while the filter moves. In this example, the value of each pixel in the output layer is the result of the dot product between the filter and the image pixels in the two squares centered on $(2, 2)$ and $(2, 4)$.}
\end{figure}

We show \ra{a sketch} of convolution in Fig.~\ref{fig:euclideanFilterOutline}: the convolutional kernel is represented by a $3\times3$ matrix, thus containing nine weights $w_{ij}$. The kernel scans the input coming from the previous layer, which, in this case, is a $3 \times 5$ pixels image. The kernel is applied to each sub-region of $3\times 3$ pixels of the input image, and the result of the dot product between the two is saved in each pixel of the output layer:
\begin{equation}
p_i = \sum_{j} w_{ij} x_{ij},
\end{equation}
where indices $x_{ij}$ are the elements of the $3\times3$ portion of the input image. The number of  pixels in the output depends on the dimension of the input, the dimension of the kernel, and the way the kernel moves across the input image (see caption of Fig.~\ref{fig:euclideanFilterOutline}). Once the output is computed, the non-linear activation function is applied to its elements, and this becomes the input of the following layer. In CNNs, training is done as for the fully connected ones, by minimizing a cost function through the application of the network on a training set.

In Fig.~\ref{fig:euclideanFilterOutline}, we applied the convolutional layer to a 2D image. Nevertheless, CNN can be applied to inputs with other dimensionality, such as 3D images (where, typically,  the third dimension deals with the color channels) or to 1D vectors. In the latter case, the kernel is a 1D vector as well, scanning the input vector and applying the dot product on a portion of it.

The fact that CNNs retain the local information of the inputs while performing the convolution on adjacent pixels makes them an extremely powerful tool for pattern recognition. In classification and regression problems,  CNNs are often used to extract synthetic information from input data. Typically, this requires some operation that reduces the dimensionality of the input, which is usually done using \emph{pooling layers}. A pooling layer partitions the input coming from the previous layer into sets of contiguous pixels and applies some operation on each set.  Therefore, a pooling layer performs some down-sampling operation; typical examples are: (1) the computation of the average over the pixels within the partition (\emph{average pooling}), or (2) the extraction of the pixel with the minimum/maximum value in the partition (\emph{min} or \emph{max pooling}).

%% file: section3.tex
\section{CNNs on the HEALPix sphere}
\label{sec:HealpixNN}

The CNN concepts and examples introduced in Sect.~\ref{sec:NN} are typically applied to Euclidean domains (e.g., 1D vectors and planar images). Nevertheless, it is of great interest to extend the method to signals projected on the sphere. This would have general applications\footnote{One example is the use of computer vision applied to images taken using $4\pi$ cameras mounted on flying drones.} and would also be extremely interesting  for those astrophysical and cosmological fields that need to study large parts of the celestial sphere.


In this work, we present a novel pixel-based algorithm to build spherical CNNs, having in mind applications related to cosmological studies.  
Our implementation relies on the HEALPix (\emph{Hierarchical Equal Area iso-Latitude Pixelization}) tessellation scheme, which we briefly introduce in the next section.

\subsection{The HEALPix spherical projection}

The HEALPix tessellation of the sphere was presented by \citet{Gorski2005}. HEALPix partitions off the spherical surface into 12 base-resolution pixels, and it recursively divides each of them into 4 sub-pixels till the desired resolution is reached. The number of time a sub-division is applied to the base pixels is equal to $\log_2(N_\text{side})$, where the parameter $N_\text{side}$ is linked to the resolution of the map. The value of $N_\text{side}$ must be a power of 2, and the number of pixels on the map is $N_\text{pix}=12\times N_\text{side}^2$.

A signal projected on the sphere  with the HEALPix scheme is usually encoded in the computer's memory as a vector, with each element being associated to a single pixel on the map, following a specific order. Two ordering schemes exist; (a) \emph{ring} order sorts pixels according to iso-latitude rings, and (b) \emph{nested} order makes use of the hierarchical structure of the tessellation to enumerate pixels. Once the ordering scheme is picked, and for a fixed value of the $N_\text{side}$ parameter, the association between the vector elements and the map pixels is uniquely defined. For a complete description of the HEALPix features and related libraries, we refer to \url{https://healpix.sourceforge.io}.

HEALPix has been widely used, with applications not only in multiple branches of astronomy and astrophysics (e.g. analysis of CMB maps \citep{planck.2015.01.mission.paper}, high-energy sources \citep{FermiSourceCatalog2015}, large-scale structure \citep{SDSS.BAO.2007,SDSS.BAO.2012}, astroparticle physics \citep{IceCube.Neutrinos.2014}, exoplanetary searches \citep{Robinson.earth.model.2011,Majeau.infrared.map.planet.2012,TESS.simulations.2015}, astrohydrodynamics \citep{ENZO.2014}, gravitational-wave astronomy \citep{gw.position.reconstruction.2016}), etc.) but also in other fields, such as the geophysical one \citep{westerteiger_et_al:OASIcs:2012:3738, Shahram07}. Algorithms able to apply CNNs on spherical maps projected with the HEALPix scheme, as the one presented in this work, have therefore the potential of being of large interest, with possible applications in several different contexts.

\subsection{Description of the HEALPix CNN algorithm}

\begin{figure}
	\centering
	\includegraphics[width=0.65\columnwidth]{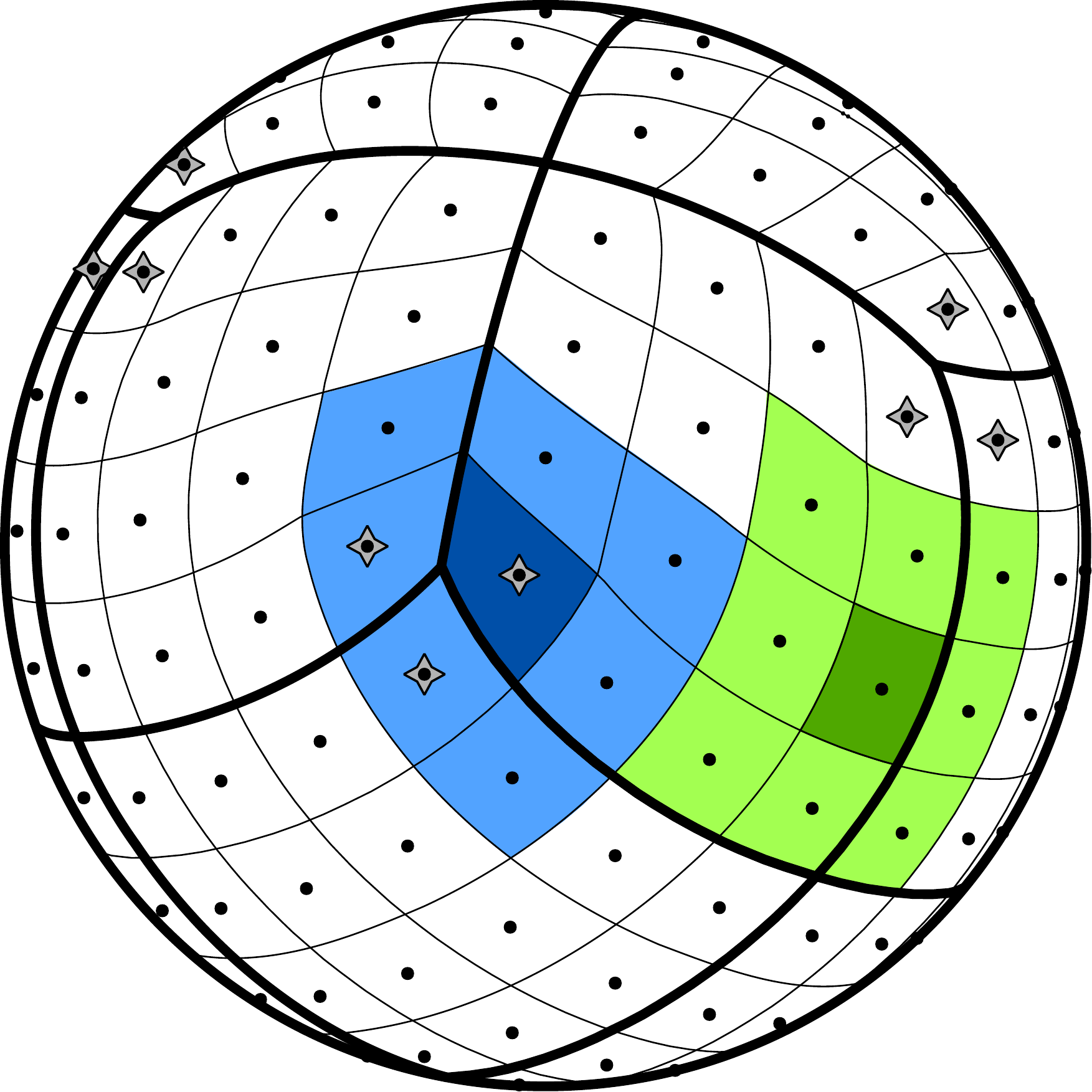}
	\caption{\label{fig:pixelNeighbors} Tessellation of a sphere according the the HEALPix scheme. In this figure the pixel resolution corresponds to $N_\text{side}=4$ ($N_\text{pix}=192$). Almost every pixel in a HEALPix map has 8 neighbors; however, a few pixels only have 7 of them. In this figure, the blue pixel has 7 neighbors, while the green one has 8. Black dots represent pixel centers; seven-neighbor pixels are highlighted with grey dimonds. These pixels are at the intersection of three base-pixels (delimited using thick lines); this kind of intersection occurs only near the two polar caps, and in any map with $N_\text{side}\geq2$ there are exactly 8 of them. Thus, the number of seven-neighbor pixels is always 24 (because of the nature of the orthographic projection used in this sketch, only 9 seven-neighbor pixels are visible).}
\end{figure}

The algorithm that we propose implements CNNs over HEALPix maps preserving the local information of the signal by applying kernels to pixels close to each other. We describe the details of this process in the following sections.

\subsubsection{Convolutional layers on a sphere}
\label{sec:convolutionSphere}

\begin{figure}
	\centering
	\includegraphics[width=\columnwidth]{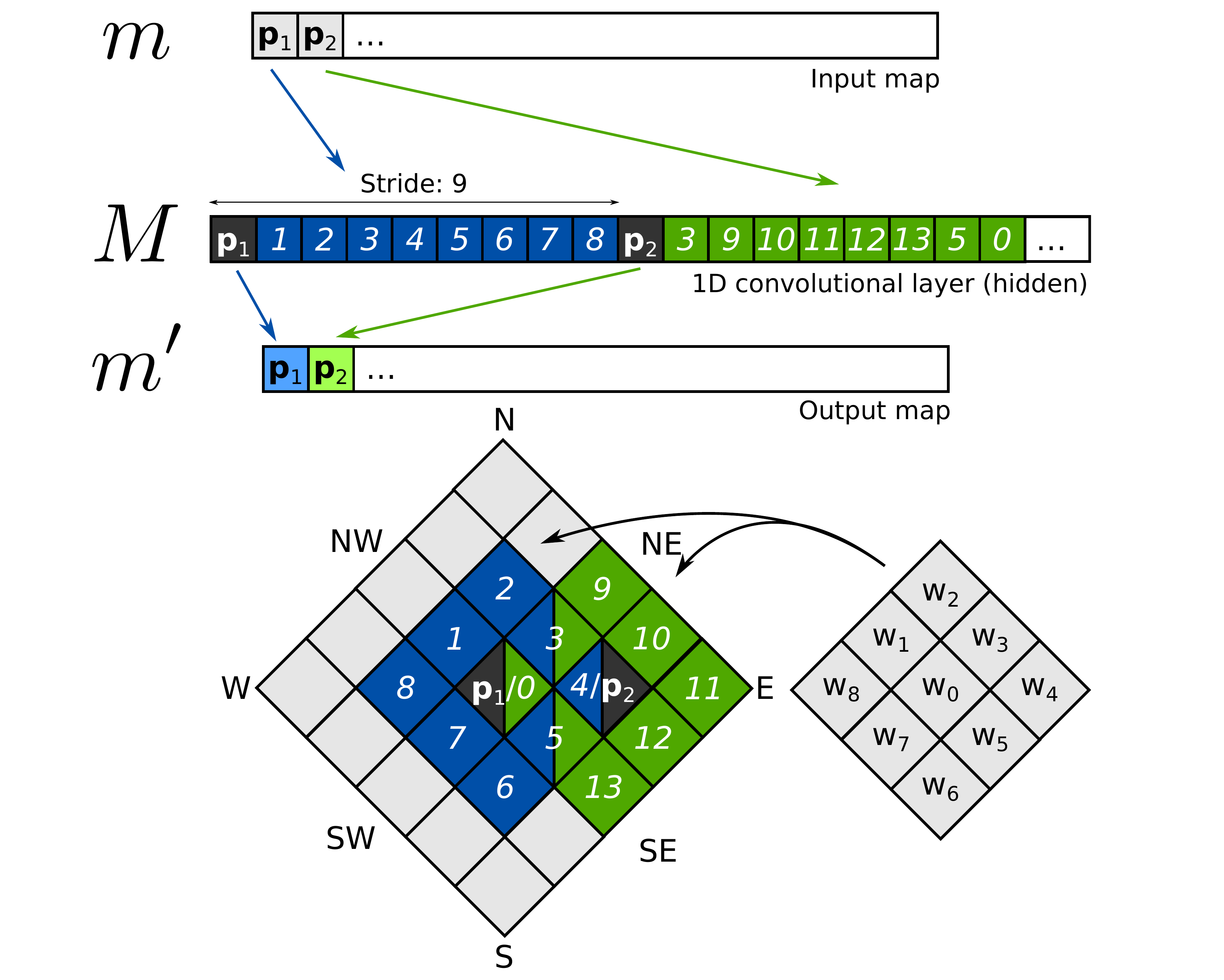}
	\caption{\label{fig:filterOutline} Sketch of the convolution algorithm developed in this work. The elements $p_1$ and $p_2$ of the vector $m$ correspond to two pixels on map (in this case $0$ and $4$). $p1$ and $p2$ are associated to nine elements each, in vector $M$. The first of this nine elements corresponds to the pixel itself, while the subsequent ones corresponds to its eight neighbors. The neighbors are visited, and unrolled in the $M$ vector, following a clockwise order staring from the NW direction. The filter (in grey) is also unrolled in a nine element long vector, using the same convention. Convolution is then implemented as a standard 1D convolutional layer with a stride of 9, and the output is the $m'$ convolved vector (map), which has the same number of elements (pixels) as the input $m$. In the figure, numbers over the pixels do not represent the values of the pixels but their unique index: in practical applications, these indices will be those associated to an ordering scheme and a map resolution $N_\text{side}$. Colors are the same as in Fig.~\protect\ref{fig:euclideanFilterOutline}, to underline the similiarity.}
\end{figure}

The majority of pixels in the HEALPix scheme has eight adjacent neighbors (structured with a ``diamond'' shape). Standard HEALPix libraries implement algorithms to find and enumerate them. In particular, for a given pixel labeled by the index $p$, these functions return the indices of its eight closest neighbors following some pre-defined ordering (N-W, N, N-E, E, S-E, S, S-W, and W). However, a few pixels (24 for $N_\text{side} \geq 2$) have only 7 neighbors: this happens for those pixels along some of the borders of the 12 base-resolution ones, as shown in Fig.~\ref{fig:pixelNeighbors}.

Our convolutional algorithm makes use of the possibility to quickly find neighbor pixels in the following way  (see Fig.~\ref{fig:filterOutline}):
\begin{enumerate}
\item the input of the convolutional layer is a map (or a set of maps) at a given $N_\text{side}$, which is stored in a vector $m$ of $N_{pix}$ elements.
\item Before applying the convolution, the vector $m$ is replaced by a new vector $M$ with $9\times N_{pix}$ elements. We associate each element $p$ in $m$ with nine elements $P^p_i$ of $M$ ($i=0\ldots 8$), setting $P^p_0$ equal to the value of the pixel $p$ itself and the others equal to the values of the eight neighbors of $p$, following the clockwise order specified above. Pixels with 7 neighbors are still associated with 9 elements of $M$, but the element corresponding to the missing pixel is set to zero.
\item Once the vector $M$ is generated, a 1D convolution is done in the standard way: a nine element long 1D kernel, with elements $w_i$,  is applied to $M$. This convolution is done with a stride parameter equal to 9, meaning that the kernel is not applied to all the elements of the $M$ vector, but it makes a jump of nine of them at each move. This is equivalent to the convolution of each pixel with its 8 closest neighbors. The elements $w_i$ of the kernel represent the weights that the network needs to optimize during the training phase.

\item The output of this operation is a new vector $m'$ with $N_{pix}$ elements, corresponding to the convolved map.
\end{enumerate} 
\ra{To visualize how convolution works we have prepared a simple example described in Appendix~\ref{sec:exampleStripes}.}

The algorithm presented above has a few interesting properties. Firstly, the search of pixel neighbors needs to be done only once for each $N_\text{side}$ value: the re-ordering scheme that brings a map $m$ into $M$ is always the same once $N_\text{side}$ is fixed, and re-ordering operation of a vector is computationally extremely fast. Secondly, the convolution is pixel based, therefore there is an immediate analogy between standard Euclidean convolution and this implementation of a spherical one. Lastly, since the operation is the 1D convolution between a kernel and a vector, it can be performed using the \ra{highly-optimized} NN libraries commonly available today (e.g., TensorFlow, PyTorch, Theano, Flux, etc.) 

\ra{On the other hand, our algorithm suffers from one main downside: unlike other spherical CNNs recently introduced in literature \citep[e.g.][]{Cohen2018,Perraudin2018}, we lose the notation of rotation equivariance. This makes our network less efficient, meaning that it needs a larger set of data to be properly trained and that its final accuracy could be sub-optimal. We discuss this point in Section~\ref{sec:related_works}.}

\subsubsection{Pooling layers on a sphere}
\label{sec:poolingSphere}

\begin{figure}[!t]
	\centering
	\includegraphics[width=0.8\columnwidth]{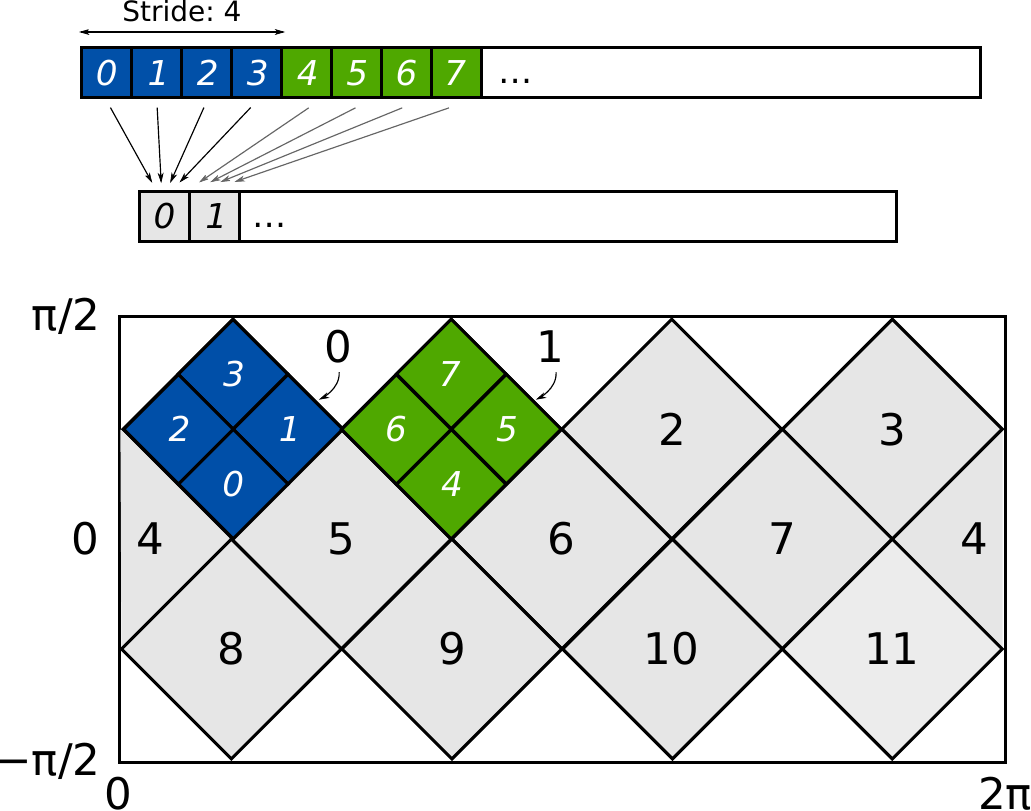}
	\caption{\label{fig:dgradeOutline} Sketch for the implementation of pooling layers on HEALPix maps. Since a map with $N_\text{side}> 1$ can be degraded to a map with $N_\text{side}/2$, thanks to HEALPix hierarchical construction, pooling can be implemented naturally using standard 1D pooling layers, if pixels of the input vector (map) $m$ are properly ordered and a stride equal to 4 is used.}
\end{figure}

As we have seen for standard CNNs (Section~\ref{sec:CNN}), pooling layers are useful for lowering the resolution of the network inputs. The hierarchical nature of the HEALPix scheme naturally supports pooling operations, as each pixel in a low-resolution map is mapped to four (or more) sub-pixels in a higher resolution map. In our algorithm, pooling is easily implemented in one dimension using the same approach described in Sect.~\ref{sec:convolutionSphere} for convolution. For each pooling layer, the input vector $m$ containing a map with $N_{pix}$ pixels is converted to HEALPix nested ordering, and a standard 1D pooling (either min, max, or average) is applied with a stride of 4. This process turns a map with resolution $N_\text{side}$ to a map with $N_\text{side}/2$. Fig.~\ref{fig:dgradeOutline} sketches the process.

\ra{\subsubsection{Related works on spherical CNNs}}
\label{sec:related_works}
\ra{In the last few years, several works have studied the possibility to apply CNNs to extract informations from signals projected on the sphere.} \ra{The simplest idea for achieving the goal relies on using standard 2D Euclidean CNNs on planar projection of the sphere, such as the equirectangular one \citep[e.g.,][]{HuLinCVPR17,Boomsma2017}.  These approaches are easy to implement, but they suffer from the fact that any planar projection of the sphere causes distortions. This problem has been addressed by the CNN implementation described in \citet{Coors2018ECCV} where the sampling location of the convolutional filters is adapted in order to compensate distortions.} 
\ra{A second approach described in literature requires to project the spherical signal to tangent planes around the sphere, and then to apply standard 2D CNNs to each of these planes \citep[e.g.,][]{Xiao12}. In the limit of making the projection on every tangent plane, this strategy is accurate, but requires large computational resources. \citet{NIPS2017_6656} proposed a combination of the two approaches by processing spherical signals in their equirectangular projection and mimicking the filter response as if they were applied to the tangent planes.}


\ra{All the methods described above propose CNNs that are not \emph{rotationally equivariant}. Rotational equivariance means that if $R \in \text{SO}(3)$ is a rotation and $f$ is the convolution operation applied to some input $x$, its output $y = f(x)$ is such that $f(R(x))=R(y)\,,\forall R$. A non-equivariant CNN architecture causes the learning process to be sub-optimal, as the efficiency in filter weight sharing is reduced.}

\ra{\citet{Cohen2018} proposes the first spherical CNN with the property of being rotational equivariant. Its approach uses a Generalized Fast Fourier Transform (FFT) algorithm to implement convolutional layers in harmonic domain. Their network is very accurate in 3D model recognition; however, it is computationally expensive, as the computation of spherical harmonics is a $O(N_{pix}^2)$ operation.}

\ra{While we were in implementing the spherical CNN described in this paper, \citet{Perraudin2018} presented an algorithm to apply the HEALPix tessellation of the sphere to CNNs. Their implementation, called \emph{DeepSphere}, uses graphs to perform convolutions. It has the advantage of being computationally efficient, with a computational complexity of $O(N_{pix}$) and to be adaptable to the analysis of signals that only partially cover the spherical surface. Moreover, by restricting to filters with radial symmetry, their network is close to be rotational equivariant.} 

Our CNN is specifically built to run on HEALPix, without requiring any other projection of sphere. Similarly to DeepShpere, it is computational efficient, scaling as $O(N_{pix}$), and it is applicable to partially-covered spherical surfaces. Moreover, since the convolution operation ends up in being a traditional 1D convolution, our algorithm has the advantage of being fully integrated in exiting libraries for NNs, like TensorFlow, and can therefore rely on years of algorithmic optimization of the convolution in the context of deep learning. On the other hand, the main drawback is the lack of rotation equivariance, which is due to three factors. Firstly, although covering all the same area, pixels in the HEALPix tessellation have different shapes, depending on their position on the sphere (see Fig.~\ref{fig:pixelNeighbors}). This causes some degree of distortion, especially close to the poles. Secondly, as described in Sect.~\ref{sec:convolutionSphere} and Fig.~\ref{fig:pixelNeighbors}, in every HEALPix sphere there are 24 pixels with only seven neighbors instead of eight. This causes distortions that mostly affects convolution on low resolution maps (i.e., small values of $N_{side}$). Lastly, in this first implementation of our algorithm we do not force filters to have radial symmetry. \ra{This makes the efficiency of our network sub-optimal. However, the good performance of our algorithm in the applications that we present in Sect.~\ref{sec:MNIST} and \ref{sec:CMBstuff} shows the validity of our approach in the present case.} 
	
\ra{It would be interesting to compare our results with the ones of other networks that encode  equivariance. However, the spherical CNN introduced by \citet{Cohen2018} is not developed to run on HEALPix; adapting the code would require an amount of work that is outside the scope of this paper. On the other hand, DeepSphere computes convolutions on the HEALPix sphere, similarly to our approach. Nevertheless, the main goal of this paper is to test the feasibility of our algortihm in estimating cosmological parameters that affect the large scales of CMB polarized maps (tensor fields), as described in Section~\ref{sec:tau}. DeepSphere has not been tested on this kind of analysis yet, as \citet{Perraudin2018} only applied their code to a classification problem performed on scalar fields. Therefore, we defer this comparison to some future works.}

%% file: section4.tex
\section{Working environment and NN architecture}
\label{sec:env}

In this section we briefly describe the working environment used to develop and train the NNs that implement the spherical convolutional layers introduced previously. We also present the network architecture used in the applications described in Section~\ref{sec:MNIST} and~\ref{sec:CMBstuff}.

\subsection{Working environment}

We have implemented the algorithm to apply CNNs on the HEALPix sphere in the \textit{Python} programming language. We have taken advantage of the existing \textit{healpy} package\footnote{\url{https://healpy.readthedocs.io/en/latest/index.html}}, a Python library specifically built to handle HEALPix maps.

As we emphasized in Section~\ref{sec:HealpixNN}, our implementation of convolutional and pooling layers ends up in being a sequence of operations on 1D vectors, which can applied using existing NN libraries. In our work, we have employed the \textit{Keras} package\footnote{\url{https://keras.io}}, with \textit{TensorFlow} backend\footnote{\url{https://www.tensorflow.org}} to build, test, and train CNNs.

We have used the computing facilities provided by the \emph{US National Energy Resource Scientific Computing Center} (NERSC) to perform training. As NERSC does not provide GPU computing nodes at the moment, we relied on CPU-based training architectures. In order to parallelize the computation and make the training more efficient, we used the \textit{Horovod} package \citep{sergeev2018horovod} to distribute the workload over cluster nodes using MPI\footnote{For a tutorial on how to work with this environment at NERSC see \url{https://docs.nersc.gov/analytics/machinelearning/tensorflow/} .}.

\subsection{Network architecture}
\label{sec:NNarchitecture}

Using the convolutional and pooling layers described in Sections~\ref{sec:convolutionSphere} and~\ref{sec:poolingSphere}, we have built a full deep CNN that we have applied to the examples reported in the following sections. For all our applications we have used a single network architecture, changing only the output layer depending on the problem addressed.

\begin{figure}
	\centering
	\includegraphics[width=\columnwidth]{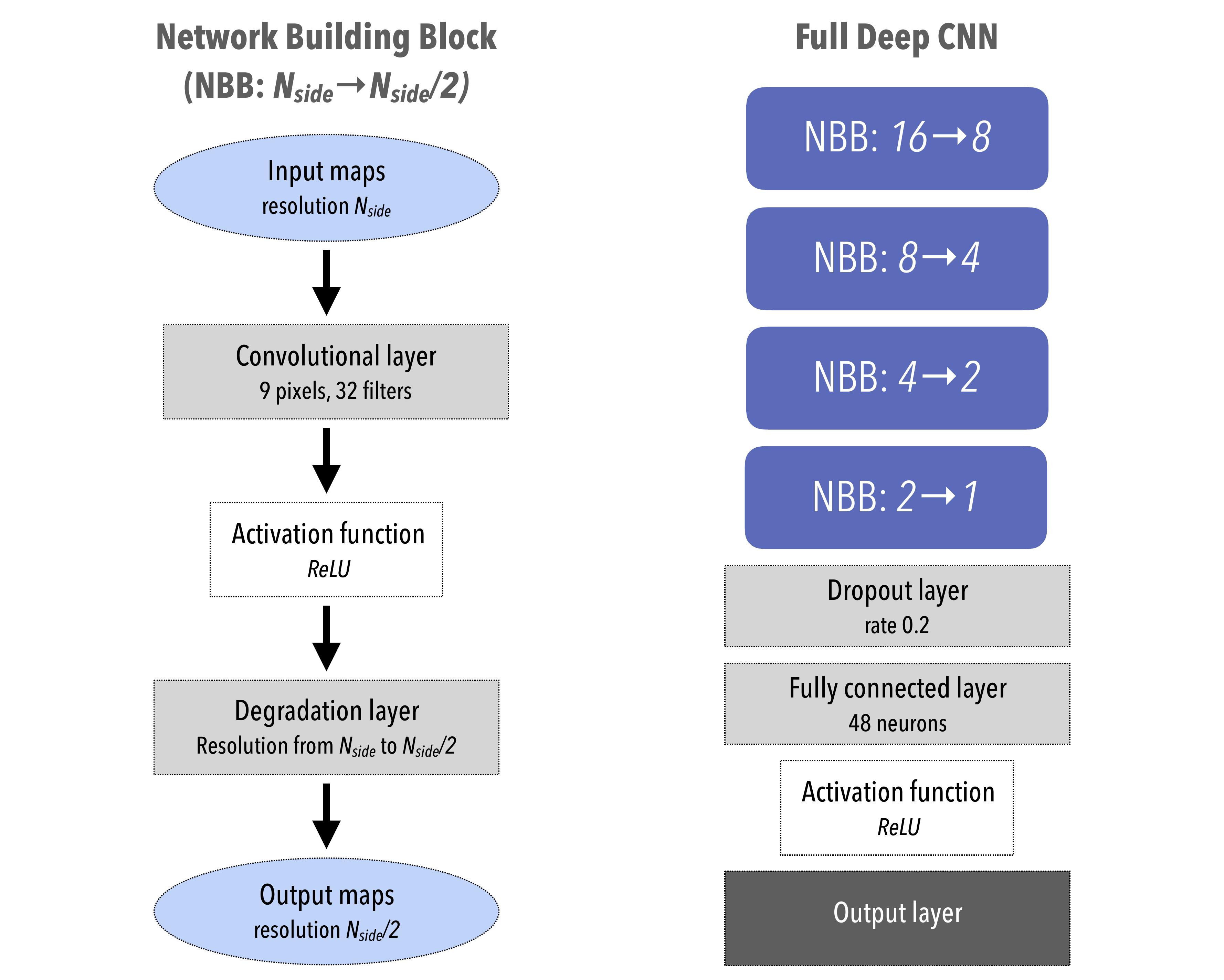}
	\caption{\label{fig:CNNarch} Left panel: basic Network Building Block (NBB). Right panel: full deep CNN architecture.}
\end{figure}

We have implemented this CNN architecture by stacking several instances of a \emph{Network Building Block} (NBB) with the following structure (see left panel of Fig.~\ref{fig:CNNarch}): 
\begin{enumerate}
\item the first layer in the NBB accepts one or more HEALPix maps with the same resolution parameter $N_\text{side}$ and performs a convolution with $N$ filters, according to the algorithm described in Sec.~\ref{sec:convolutionSphere} and Fig.~\ref{fig:filterOutline}, then applying an activation function;
\item the next layer contains an average-pooling operation (see Sec.~\ref{sec:poolingSphere} and Fig.~\ref{fig:dgradeOutline});
\item as a result, the NBB lowers the resolution of the input maps from $N_\text{side}$ to $N_\text{side}/2$.
\end{enumerate}
\par
In all our applications, the input of the network is a set of HEALPix maps at $N_\text{side}=16$ (each map has $3072$ pixels, each of which covers an area of about $13\,\text{deg}^2$ on the sphere). We have built the full deep CNN by stacking together four NBBs, going from maps at $N_\text{side}=16$ to $N_\text{side}=1$. In each convolutional layer, we convolve the input maps with $N=32$ filters. We use a \emph{rectified linear unit} (ReLU) as the activation function in the NBB.

After the four building blocks, we include a \emph{dropout} layer, with a dropout rate set to 0.2, in order to avoid overfitting\footnote{The dropout regularization randomly drops neurons and their connections in a NN layer during training. The dropout rate represents the percentage of the neurons of the layers randomly switched off in each training epoch.}. Subsequently, the 32 output maps are flattened into one vector, which is fed to a fully connected layer with 48 neurons activated via a ReLU function. The shape of the output layer and its activation function depend on the kind of problem considered. The full CNN architecture is shown on the right panel of Fig.~\ref{fig:CNNarch}.

This architecture, which convolves maps at decreasing angular resolution while moving forward in the network, makes each layer sensitive to features on map laying at different angular scales. As it will be described in the following, this is an important property for the proposed applications.

%% file: section5.tex
\section{Application to the MNIST dataset}
\label{sec:MNIST}

The first application of the algorithms presented in this work is the classification of handwritten digits projected over the sphere. We have used images from the MINST image database\footnote{The MNIST (Modified National Institute of Standards and Technology) database can be downloaded from \url{http://yann.lecun.com/exdb/mnist/}}, a publicly available database containing 70,000 grayscale $28\times28$  pixels images with handwritten digits (having a repartition between training and test set with a proportion $6:1$).\par

We have chosen this example, despite being of little relevance for astrophysics, because the MNIST dataset is widely used to test the performance of machine learning algorithms, and it is currently considered one of the standard tests to check how well automatic classifiers perform. Moreover, the performance of CNNs depends critically on the existence of a large and reliable training set, and MNIST satisfies this requirement.

\subsection{Projection}

\begin{figure}[!h]
	\centering
	\includegraphics[width=0.4\columnwidth]{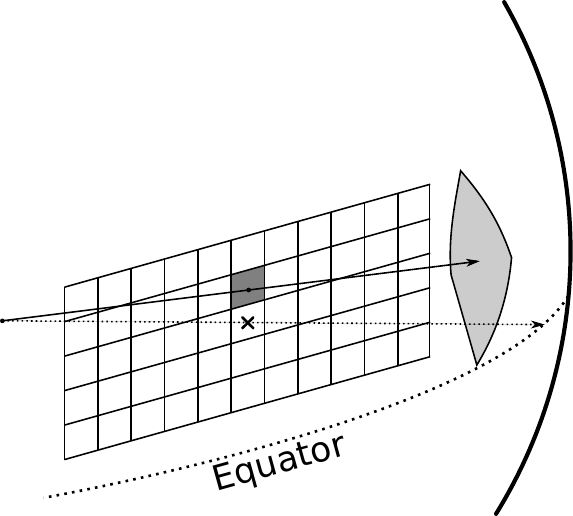}
	\caption{\label{fig:raytracing} Method used to project MNIST images on the sphere. The image is aligned so that the center of the pixel in the middle (marked with a $\times$) is on the Equatorial plane. A number of rays are fired, originating from the center of the sphere, targeting each of the pixels in the image, and the intersection with each pixel on the sphere is computed. In order to cover all the pixels on the sphere, the angular separation among rays is $1/2$ of the angular resolution of the HEALPix map}
\end{figure}

To use MNIST images in our tests we need to project them on the HEALPix maps. We have therefore developed a simple algorithm to project a rectangular image over a portion of the sphere, as shown in Fig.~\ref{fig:raytracing}. We consider the image as placed on a plane in 3D space, perpendicular to the Equatorial plane of the celestial sphere; the point at the center of the image is on the Equatorial plane. We use a ray tracing technique, firing rays from the center of the  sphere against pixels in the plane and finding their intersection with the  sphere. We fill each pixel on the HEALPix map hit by a ray with the value of the corresponding pixel in the plane figure. We fire a number of rays sufficient to cover all the HEALPix pixels in the portion of the sphere where we want to project the image, and we bin multiple hits that fall in the same spherical pixel with an average. In this way, the image on the sphere is smooth and has no holes. After the image has been projected with its center on the Equator, the code applies a rotation, in order to center the image on a user-defined point on the sphere.

\subsection{Training and results}

We have performed the training of the NN on the Cori cluster at \emph{NERSC}, with the working environment and network architecture described in Section~\ref{sec:env}. In particular, since this application of our algorithm is a classification problem the last layer of the CNN contains 9 densely-connected neurons, corresponding to the nine different classes of our problem\footnote{Once projected on the sphere with random rotation, numbers 6 and 9 cannot be distinguished. Therefore, we drop out all images containing digit 9 from the MNIST dataset and only use nine classes, corresponding to integer numbers from 0 to 8.}. We have used a \textit{Softmax} activation function for the last layer, as this is the typical choice for classifiers.

To generate the training set, we have taken 40,000 images out of the MNIST database, and we have projected them on the HEALPix sphere, randomizing on the position, rotation, and size of each image. Each projected image has random dimensions with $120^{\circ}<\theta<180^{\circ}$ and $120^{\circ}<\varphi<360^{\circ}$, therefore covering a large portion of the sphere. We have used this basic data-augmentation to produce a training set of 100,000 different images. The validation set is composed by 10,000 randomly projected images, taken from the remaining 10,000 ones of the MNIST training set.

\begin{figure}
	\centering
	\includegraphics[width=\columnwidth]{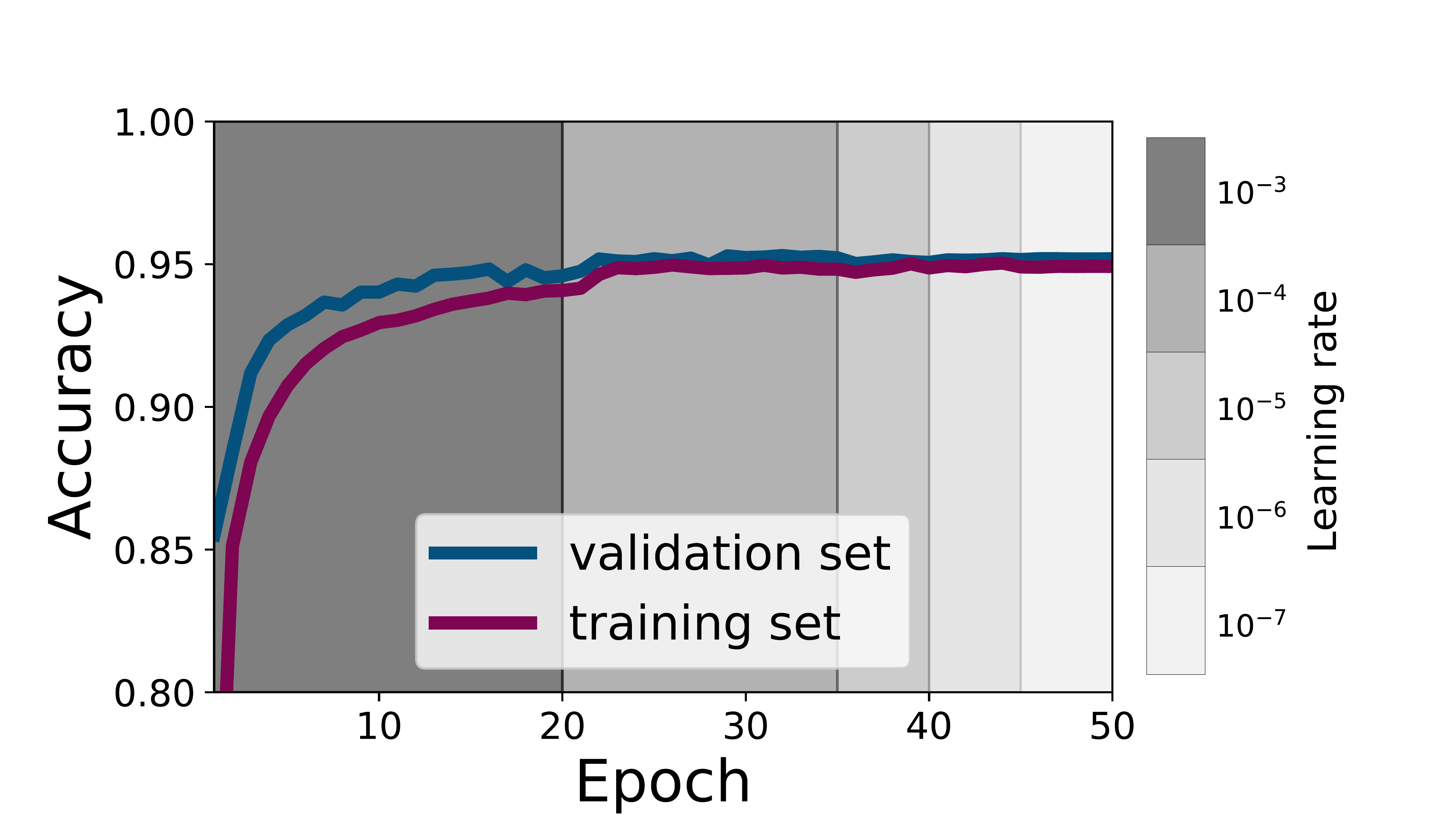}
	\caption{\label{fig:mnist_history}Learning history for the CNN applied to the MNIST dataset projected on the sphere.}
\end{figure}

We have adopted the \emph{mean squared error} as the loss function for training, and \emph{categorical accuracy} as the metric to assess the NN performance. We have used the \emph{Adam} optimizer on mini-batches of 32 samples, with  learning rate equal to $10^{-3}$ in the first step and then divided by 10 every time the validation loss has not improved for 5 consecutive epochs. We stopped the training after twenty epochs without improvement. Each epoch required $\sim 350\,\text{s}$; training stopped after $50$ epochs, for a total of less than 5~hours of wall-clock time. Figure~\ref{fig:mnist_history} reports the training history of the network. The accuracy achieved on the validation set is $\sim95\,\%$. The network did not show overfitting, but it generalized well the performance from the training set to the validation one.

We have generated 10,000 randomly projected images from the MNIST database to be used as test set, using the same data-augmentation described above. The error we achieved on the test set is about $4.8\%$.  This error is comparable to the one obtained for the classification of the MNIST dataset projected on the sphere by \citet{Cohen2018}.  

\begin{figure}[!h]
	\centering
	\includegraphics[width=\columnwidth]{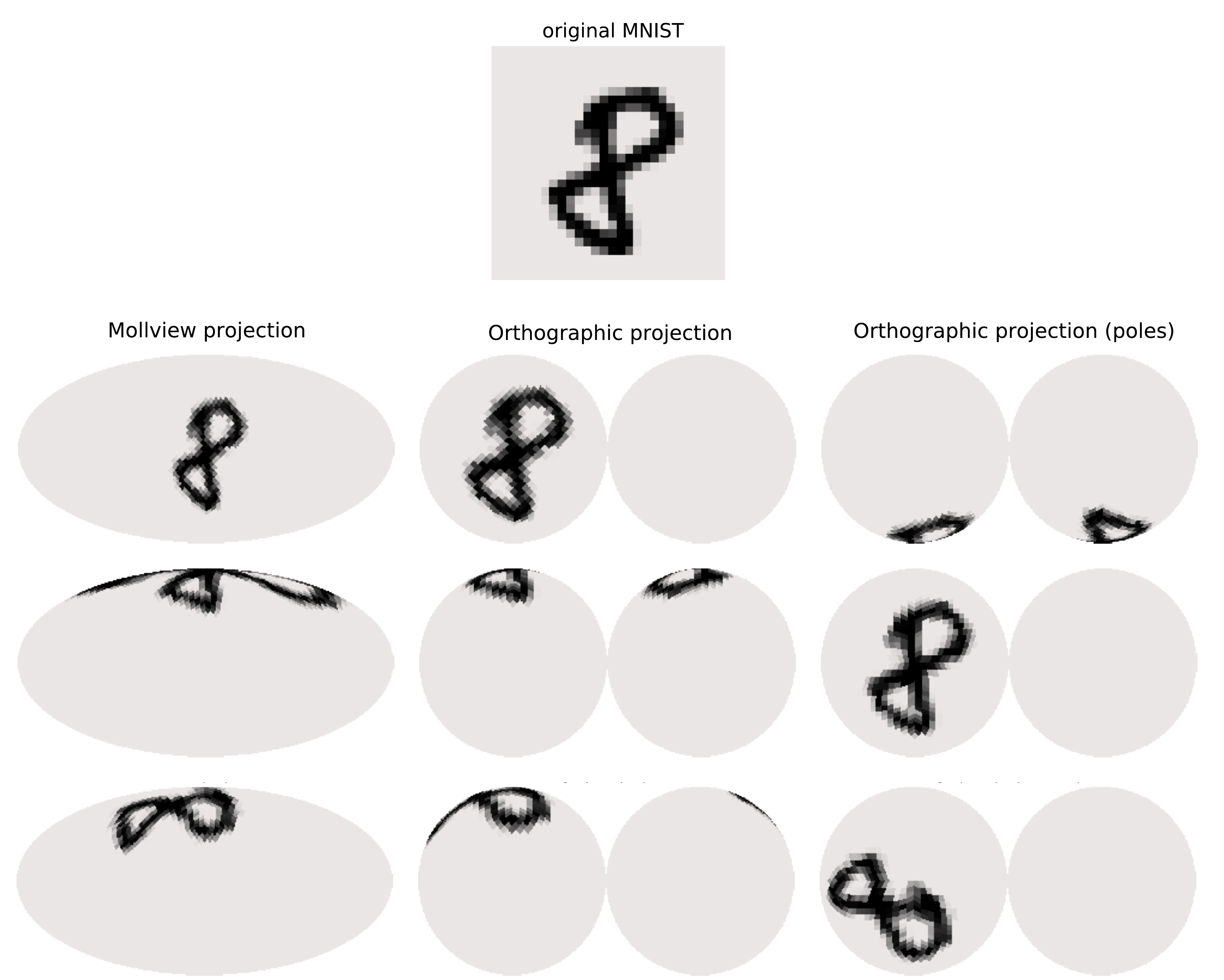}
	\caption{\label{fig:mnist_proj} Projection of a MNIST image (upper panel) on the HEALPix sphere. The same image is projected along the Equator (first row), around the north pole (second row) and at a random position on the sphere (third row). For each case, we show the map using the Mollweide projection (first column), an Orthographic projection centered on the Equator and at the north and south poles. The resolution of the projected maps corresponds to $N_\text{side}=16.$ }
\end{figure}

\begin{figure}[!h]
	\centering
	\includegraphics[width=0.9\columnwidth]{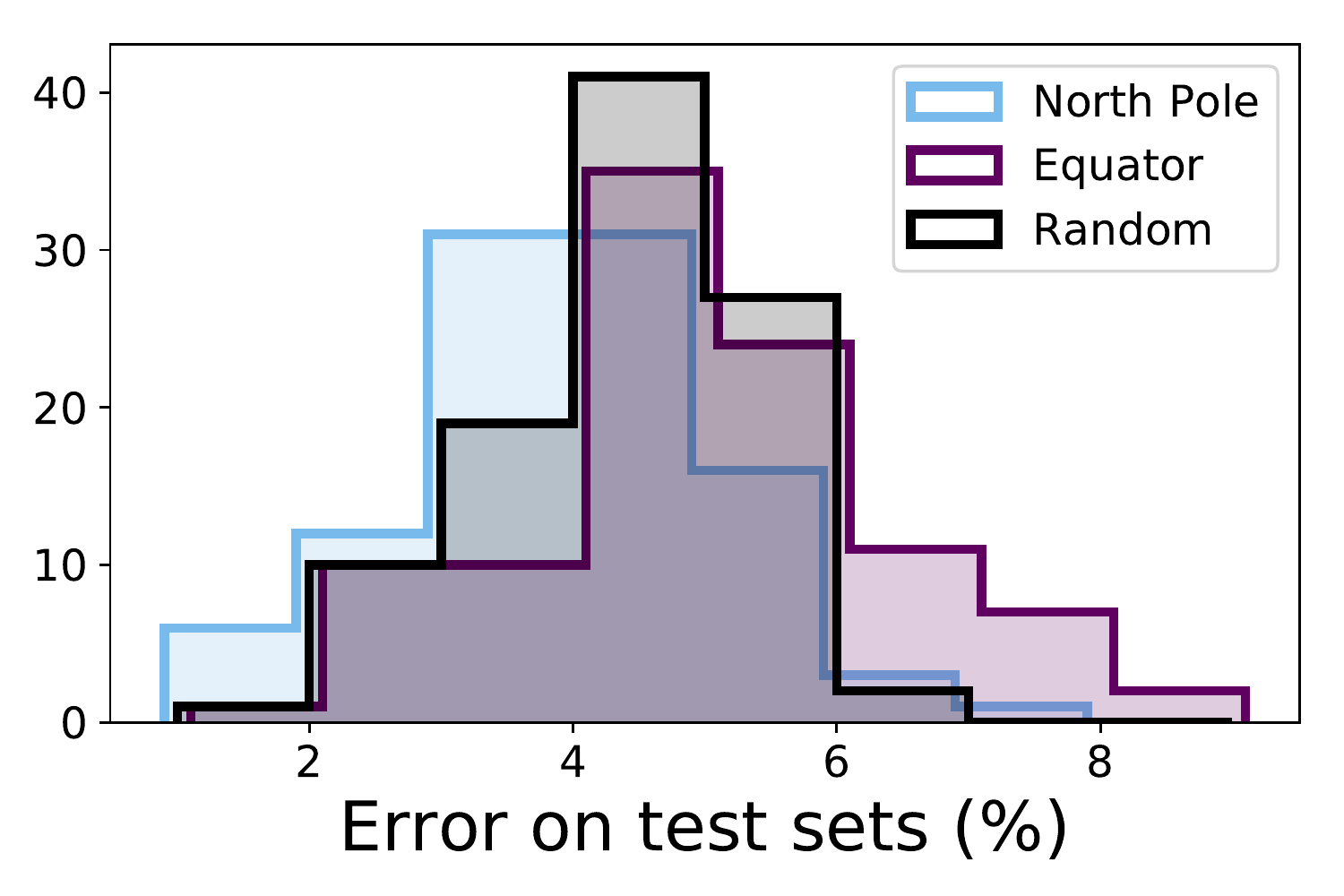}
	\caption{\label{fig:mnist_position}Distribution of NN error on three groups of test sets, where MNIST images have been projected: (i) at the north pole (blue), (ii) at the equator (purple) and (iii) on random positions on the sphere (black). Each group is composed by 100 test sets of 300 images each.}
\end{figure}

\ra{As discussed in Section~\ref{sec:related_works} our CNN is not rotationally equivariant and distortions, due to the different pixel shapes, could affect the network performance, especially close to the poles. For this reason we have checked that, once trained,} the network is able to classify handwritten digits with the same accuracy, independently from their position on the sphere. To do so, we have tested the network on three different groups of images, each containing 100 different sets of 300 images for a total of 30,000 samples for group. The three groups differed only in the region of the sphere where we projected the image:
\begin{enumerate}
	\item images in the first group were projected at random points along the Equator;
	\item images in the second group were projected near the North Pole;
	\item images in the third group were projected in random positions on the sphere.
\end{enumerate}
We show examples from each group in the first, second and third raw of Fig.~\ref{fig:mnist_proj} respectively. We report results in Fig~\ref{fig:mnist_position}: for each group, we provide the histogram of the final error for the one hundred different test sets. The three distributions are close to each other, showing that the network performs equally and independently on the position of the object on the sphere. 

%% file: section6.tex
\section{Parameter estimation for CMB}
\label{sec:CMBstuff}

As a second application of our algorithm, we have applied CNNs on the sphere to a regression problem. Our objective was to assess the performance of the NN when applied to the estimation of parameters defining the properties of a Gaussian \emph{scalar} or \emph{tensor} field on the sphere. \ra{This represents the main goal of this explorative work: to understand whether our spherical CNN implementation is suitable to be applied to this kind of problem.}

The CMB represents the relic radiation from the Big Bang, emitted when the universe was about 380,000 years old, after the decoupling of matter and photons. It has a blackbody spectrum at the temperature of $\sim2.7\,\mathrm{K}$ and it shows a high level of isotropy in the sky. Anisotropies in its temperature, which represent the trace of primordial fluctuations, are of the order of $\Delta T/T\simeq10^{-5}$. \ratwo{The CMB signal is linearly polarized due to Thomson scattering of photons with free electrons in the primordial universe. A net linear polarization can be generated only when this scattering occured in the presence of a quadrupole anisotropy in the incoming radiation. Since only a small fraction of photons lastly scattered under this condition the polarization fraction of the CMB radiation is rather small, at the level of about 10\%.} 

In the last decades, several experiments have measured the CMB anisotropies\footnote{See \url{https://lambda.gsfc.nasa.gov/product/expt/} for a list of all previous and operating experiments}, \ratwo{in both total intensity and polarization}, making the study of the CMB radiation one of the fundamental branches of modern observational cosmology.
In particular, information on the origin, evolution and composition of our Universe can be extracted with a statistical analysis of the CMB temperature and polarization fields \ratwo{used to estimate the values of the cosmological parameters). As primordial CMB fluctuations can be considered as a Gaussian field all the information is encoded in their angular power spectrum.} Classical methods for parameter fitting rely on Bayesian statistics: the values of cosmological parameters, defining the properties of our Universe, are typically extracted through the maximization of a likelihood function computed from the angular power spectra of CMB sky maps.

Recently, several works have explored the possibility to estimate cosmological parameters directly from CMB maps using NNs  without the computation of angular power spectra. \citet{he2018analysis} have successfully used deep residual CNNs to estimate the value of two cosmological parameters from CMB temperature sky maps, in the flat sky approximation. \ra{the DeepSphere algorithm presented by \citet{Perraudin2018} have been tested to classify weak lensing convergence maps (which are scalar fields similarly to CMB temperature maps) generated from two sets of different cosmological parameters.} \citet{2018arXiv181001483C} also have applied CNNs to the field of CMB science, extracting the projected gravitational potential from simulated CMB maps whose signal is distorted by the presence of gravitational lensing effect.

As pointed out by the authors of these papers, the possibility to complement traditional Bayesian methods for cosmological parameter estimations with innovative network-based alternatives could represent an important tool to cross-check results, especially in the case where the hypothesis of signal Gaussianity does not hold. This happens if the CMB primordial radiation is contaminated by non-Gaussian spurious signals, such as residual instrumental systematic effects or Galactic emissions.

In the following sections, we demonstrate that the algorithm presented in this work has the potentiality to be used for this purpose. We have first applied the NN to a ``mock'' cosmological parameter, in order to asses its ability to \ra{perform regression and} estimate it directly from maps (Sec.~\ref{sec:gauss_par}). We have then tested the NN performance on the estimation of the value of $\tau$, \ra{one of the six fundamental cosmological parameters of the $\Lambda CDM$ model,} which quantifies the Thomson scattering optical depth at reionization, on simulated maps (Sect.~\ref{sec:tau}).

\subsection{Position of a peak in the power spectrum}
\label{sec:gauss_par}

\begin{figure}[!h]
	\centering
	\includegraphics[width=\columnwidth]{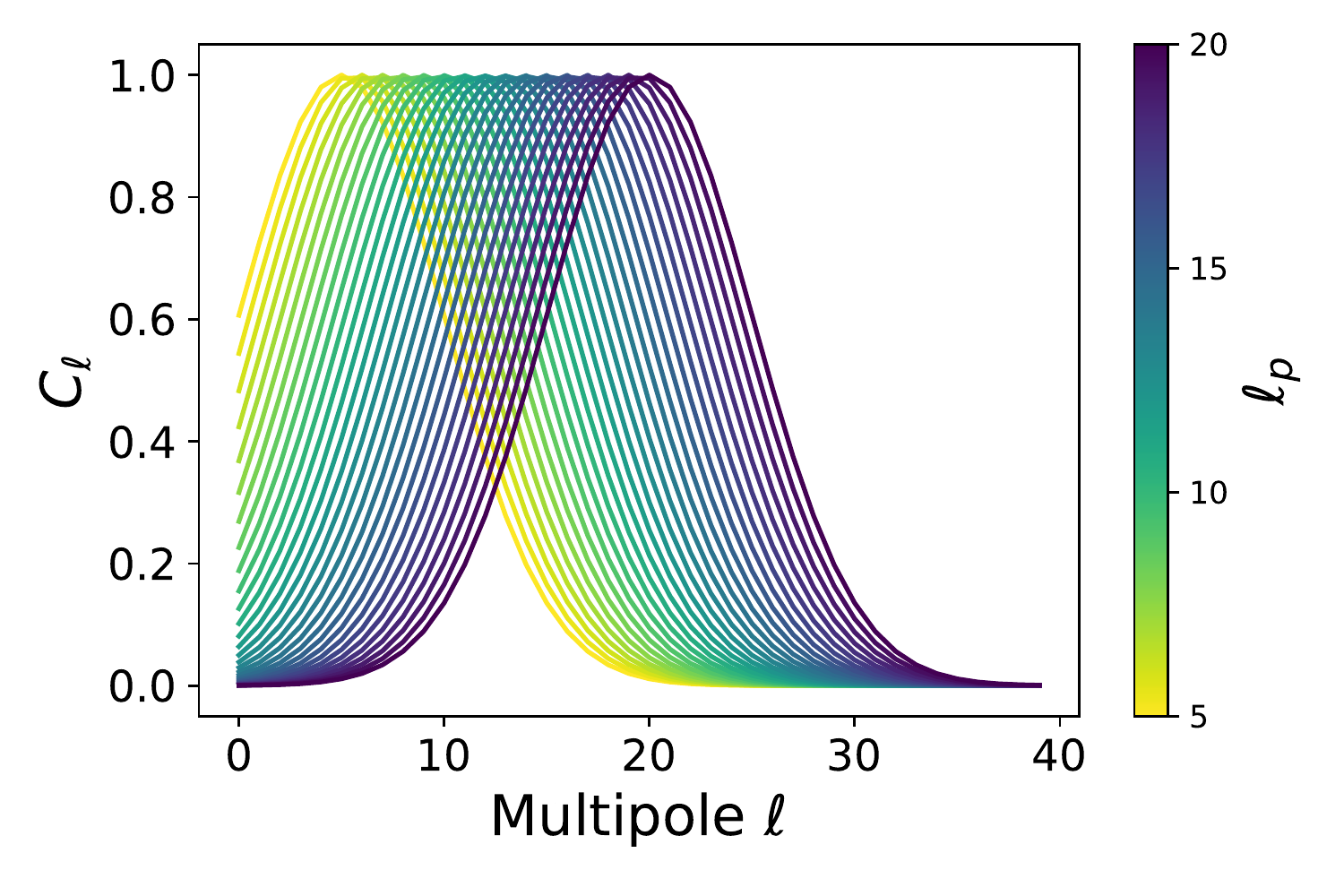}
	\caption{\label{fig:ellp_spectra}Power spectra computed according to Eq.~\ref{eq:ellp_spectra}. The $\ell_p$ parameter, defining the position of the peak, varies in the multipole range $5-20$.}
\end{figure}

We have assessed the ability of our CNN to estimate the value of a mock cosmological parameter in pixel-space. This parameter, $\ell_p$, defines the angular scale of the peak of a random Gaussian field in the power spectrum. We have constructed simulated maps starting from power spectra having the following form:
\begin{equation}
C_{\ell} = \exp\left(-\frac{(\ell-\ell_p)^2}{2\sigma_p^2}\right)+10^{-5}.
\label{eq:ellp_spectra}
\end{equation}
The peak in the power spectrum is centered around multipole $\ell_p$, with standard deviation\footnote{We have added a constant small plateau at $10^{-5}$ in order to avoid convergence issues.} $\sigma_p$. We have fixed the value of $\sigma_p=5$, while we have varied the value of the parameter $\ell_p$ in the range of multipoles between $5$ and $20$. Examples of these power spectra, for different values of $\ell_p$, are shown in Fig.~\ref{fig:ellp_spectra}.\par

To assess the feasibility of CNNs to retrieve cosmological parameters under different conditions, we have estimated the value of $\ell_p$ in the following cases: on scalar and tensor fields, with or without noise, in the case of full or partial sky coverage, as described in the following sections.

\subsubsection{Application to a scalar field}
\label{sec:applScalarField}

We have constructed the training, validation and test sets by simulating maps as random Gaussian realizations of the power spectra defined in Eq.~\ref{eq:ellp_spectra} using the \emph{synfast} module in the \emph{healpy} package. We have generated a scalar field projected on the sphere for each spectrum, corresponding to a temperature map in the CMB terminology. Maps have been simulated from spectra with $\ell_p$ randomly chosen from a uniform distribution in the interval $5-20$ at the resolution corresponding to $N_\text{side}=16$. We have not included noise in this first test.

We have used the working environment described in Section~\ref{sec:env}. We have used 100,000 maps for the training set, and 10,000 maps for the validation set; we have considered different seeds and $\ell_p$ values for each each map. The network architecture is the same as the one applied to the MNIST example (see Section~\ref{sec:NNarchitecture}), with the only exception of the last layer: as our goal was to estimate the parameter $\ell_p$, the output layer contained one single neuron.  We have trained the network using a mean squared error as  loss function, and we have monitored the accuracy of the network using the mean percentage error on the validation set. Training has been done with Adam optimizer and learning rate decay, as for the MNIST case, on mini-batches of 32 sample. The convergence took place after 70 epochs\footnote{The code stopped training after 20 epochs without improvement of the validation loss.}, for a total of about 6 wall-clock hours of training.

We have applied the trained network on a test set, composed by 1,000 maps simulated as before, and we have compared the results reached with the network with those obtained with standard Bayesian method. In the latter case, we have estimated the $\ell_p$ parameter from the power spectrum of each map of the test set (computed with the \emph{anafast} module within \emph{healpy}), using a Markov Chain Monte Carlo (MCMC) algorithm to maximize the following $\chi^2$ likelihood:
\begin{equation}
-2\ln\mathcal{L}= \sum_{\ell}{(\widehat{C}_{\ell}-\widetilde{C}_{\ell})^2/\sigma(\widetilde{C}_{\ell})^2},
\label{eq:loglike}
\end{equation}
where $\widehat{C}_{\ell}$ is derived from the reference model (in this case, $\widehat{C}_{\ell}=C_{\ell}$ from Eq.~\ref{eq:ellp_spectra}), $\widetilde{C}_{\ell}$ are computed from the simulated map and $\sigma(\widetilde{C}_{\ell})$ represent their uncertainties. Since we have been running on simulations, we assume that $\sigma(\widetilde{C}_{\ell})=\sigma(\widehat{C}_{\ell})$, with the signal variance equal to
\begin{equation}
\sigma(\widehat{C}_{\ell})=\widehat{C}_{\ell}\sqrt{\frac{2}{(2\ell+1)}}.
\label{eq:cv}
\end{equation}
We fit the power spectra up to $\ell_{max}=(3N_\text{side}-1)=47$.

The mean error obtained with CNN on the test set is $\sim1.3\,\%$, while the error of the MCMC fit is $\sim0.7\,\%$. Although the standard Bayesian method performs about twice as good as the NN, there are two important considerations to be made, which make the network performance remarkable: 
\begin{enumerate}[i.]
\item we have not optimized our network architecture, as we have used the same architecture described in Section~\ref{sec:NNarchitecture} adapted to the regression problem. Indeed, as already emphasized, the goal of our work is not to find the best architecture for a given problem, but to prove the feasibility of our approach. An optimized architecture could therefore lead to better results in the estimation of the $\ell_p$ parameter.
\item Contrary to the standard maximum likelihood approach applied to power spectra to fit for the value of $\ell_p$, the estimation with the network does not make any assumption on the Gaussianity of the signal.
\end{enumerate}

\begin{figure}
	\centering
	\includegraphics[width=\columnwidth]{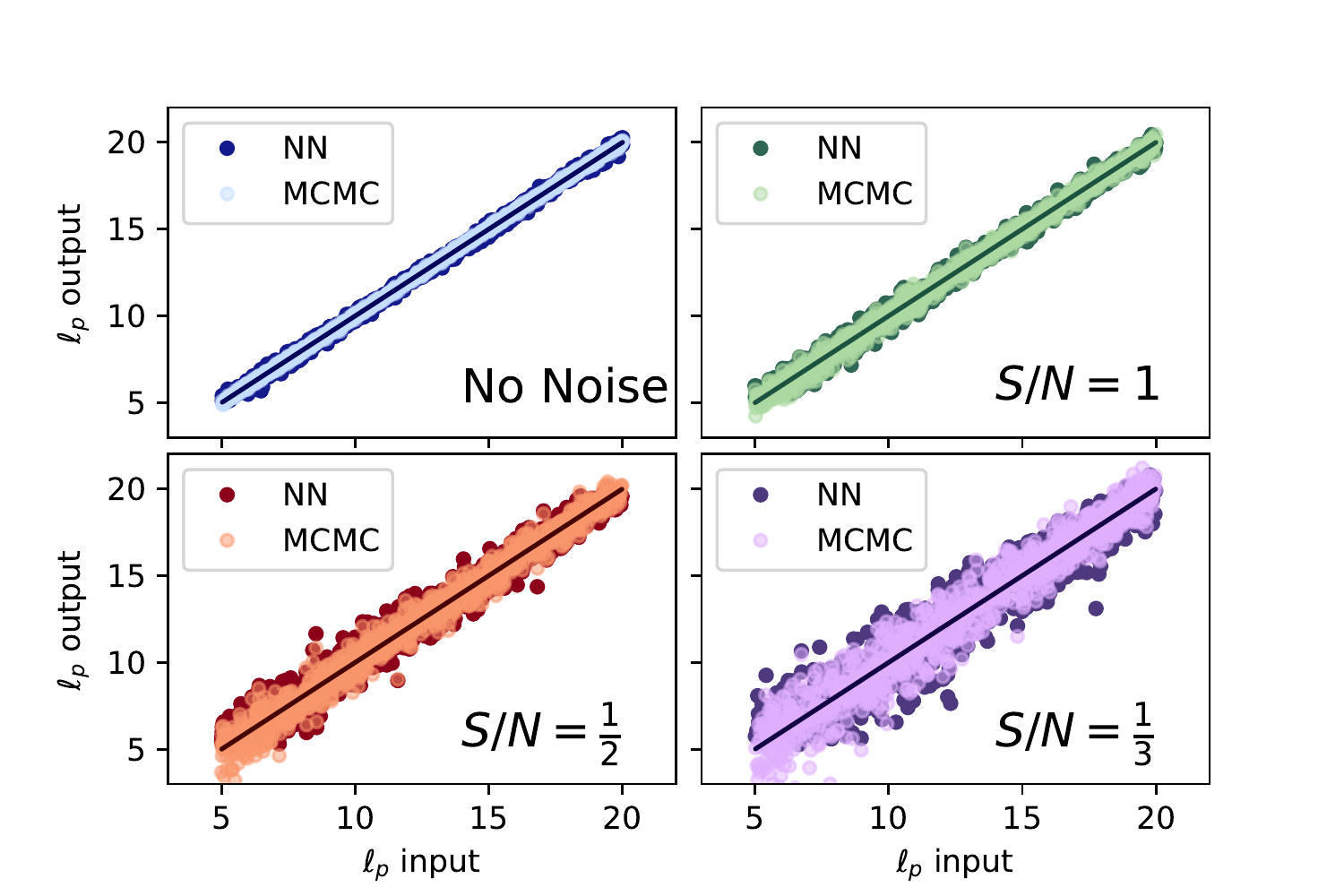}
	\caption{\label{fig:gauss_noise} Comparison of the results obtained in the estimation of the $\ell_p$ parameter with the neural network and with the MCMC fit, for the noiseless case and for the three considered noise levels.}
\end{figure}

We have also tested the network performance in the presence of white noise. We have considered three levels of noise on maps, with standard deviation $\sigma_n=5, 10, 15$, corresponding to a signal-to-noise ratio\footnote{The signal amplitude is defined as the standard deviation of the noiseless maps computed on the whole training set of 100,000 maps. This means that the signal-to-noise ratio is slightly different on each map of the training set, as the signal is generated from spectra with different value of $\ell_p$, while the noise standard deviation is the same for all maps.} of about $1$, $1/2$ and $1/3$. For each of the three noise levels, we have trained a new NN, with the same architecture defined above and a training set containing 100,000 signal+noise maps. We have generated each map as a Gaussian realization of the power spectrum of Eq.~\ref{eq:ellp_spectra} plus Gaussian noise. The training strategy has been the same used for the noiseless case.

As in the previous case, we have compared the results from CNN with the ones coming from standard MCMC fitting on a test set of 1,000 maps, for each of the three noise levels.  When fitting for the value of $\ell_p$ from the power spectra, we have taken into account the fact that the presence of noise on maps induces a bias on the auto-spectra $\widetilde{C}_{\ell}$, which must be considered while maximizing the likelihood in Eq.~\ref{eq:loglike}. Therefore, in this case, the reference model becomes:
\begin{equation}
\widehat{C}_{\ell} = C_{\ell}+N_{\ell},
\end{equation}
with $C_{\ell}$ from Eq.~\ref{eq:ellp_spectra} and $N_{\ell}$ being the noise power spectrum.  In the case of homogeneous white noise with standard deviation $\sigma_n$ on a map with $N_{pix} = 12\times N_\text{side}^2$, the noise power spectrum is:
\begin{equation}
N_{\ell} = \frac{4\pi\sigma^2_n}{N_{pix}}.
\end{equation}

We present a summary of the results obtained with the NN and with the  MCMC approach in Table~\ref{tab:gauss_noise}. For each noise level used in the simulation, we report the mean percentage error on the estimated values of the $\ell_p$ parameter computed over the entire test set. In Fig.~\ref{fig:gauss_noise}, we show the distribution of the estimated values of $\ell_p$ around the true ones for each noise level, considering both approaches. Even in the presence of noise, the accuracy is comparable with the one reached with the MCMC fit.

\begin{table}
      \caption{Summary of the results obtained in the estimation of the $\ell_p$ parameter from the neural network and the MCMC fit, for the noiseless case and the three considered noise levels. The values reported represent the mean percentage error (percentage difference between the input and the estimated $\ell_p$ value) computed over the entire test set of 1,000 samples. }
         \label{tab:gauss_noise}
     $$ 
         \begin{tabular}{lcccc} 
            \hline\hline
            \noalign{\smallskip}\noalign{\smallskip}
          & Noiseless &   $S/N=1$ &  $S/N=1/2$ & $S/N=1/3$ \\   
          &$\sigma_n=0$ & $\sigma_n=5$ & $\sigma_n=10$ & $\sigma_n=15$\\        
            \noalign{\smallskip}
             \hline
            \noalign{\smallskip}
             \noalign{\smallskip}
              NN            &1.3\,\% &2.9\,\% & 5.2\,\% & 8.4\,\%       \\
               \noalign{\smallskip}      
              MCMC      &  0.7\,\% &   2.5\,\%& 4.8\,\%&7.8\,\%   \\                             
            \noalign{\smallskip}
            \hline
         \end{tabular}
     $$ 
      \end{table}

\subsubsection{Application to a tensor field}
\label{sec:tensor_field}

We have tested the ability of the network to estimate $\ell_p$ also in the case of a tensor field projected on the sphere. This is analog to the case of CMB polarization measurements, and it is of great importance, as current CMB experiments are mainly focusing on polarization observations.

The CMB signal is linearly polarized, with a polarization fraction of about $10\,\%$, and with amplitude and orientation defined by the Stokes parameter $Q$ and $U$. Although CMB experiments directly measure the amplitude of Stokes parameters across the sky, producing $Q$ and $U$  maps, these quantities are coordinate-dependent. Therefore, it is more convenient to describe the CMB field using a different basis, dividing the polarization pattern according to its parity properties and using the so-called \emph{E-modes} (even) and \emph{B-modes} (odd).  The use of the $(E,B)$ basis is convenient also from a physical point of view, as these modes are sourced by different types of perturbations in the primordial universe, with $B$-modes produced by tensor (metric) perturbations.

Using tensor spherical harmonics, the polarization tensor field can be described in terms of $E$ and $B$ power spectra, which, as for the case of a scalar (temperature) field, give a complete description of the signal statistics in case of a Gaussian field.

For our exercise, we have used the model in Eq.~\ref{eq:ellp_spectra} to build $E$ and $B$-mode power spectra: from those, we derive Stokes $Q$ and $U$ maps with $N_\text{side}=16$. For each pair of maps we have considered polarization power spectra with values of the parameters $\ell_p^E$ and $\ell_p^B$ different from each other and randomly draw from a uniform distribution in the multipole interval $5-20$.

In this application, the purpose of the NN was to estimate the values of both  $\ell_p^E$ and $\ell_p^B$ having as input a pair of $Q$ and $U$ maps. In order to do so, we slightly change the input and output layers of the network: the input layer accepts two maps ($Q$ and $U$) instead of one, and the output layer emits two scalar values ($\ell_p^E$ and $\ell_p^B$). The other network layers share the same architecture used for the recognition of the MNIST dataset in Sect.~\ref{sec:MNIST} and for the regression problem on a scalar field in Sect.~\ref{sec:applScalarField}.

The training set was composed by 100,000 pairs of $Q$ and $U$ maps, generated from 100,000 pairs of $E$ and $B$ spectra, and the validation set contained 10,000 pairs of maps. We have trained the network using the same procedure described in Sect.~\ref{sec:applScalarField}: we have minimized a mean squared error loss function, reaching convergence on the validation set after about 7 hours of training.

As before, we have assessed the performance of the NN on a test set of 1,000 samples. The mean percentage error on the estimation of $\ell_p^E$ and $\ell_p^B$ is about $2.7\,\%$ on the test set. This number needs to be compared with the accuracy that can be reached with standard MCMC fitting. In this case we have estimated $\ell_p^{E/B}$ by computing the $E$ and $B$-mode power spectra of the test maps, and  by minimizing, separately for the two spectra, a chi squared likelihood (see Eq.~\ref{eq:loglike} and~\ref{eq:cv}) as we did for the scalar case. The mean percentage error that we get in on both $\ell_p^E$ and $\ell_p^B$ from this standard fitting procedure is about $0.7\,\%$ (analogous to the scalar case).

Although it is true that for polarization field the NN \ra{performs} worse than the standard MCMC approach, it is important to notice that, with this simple exercise, we demonstrate for the first time that the network can still discern between $E$ and $B$-modes and fit for two parameters that independently affect the statistics of the two polarization states. Also, as already emphasized, we have not optimized the network architecture, but have stuck to the one used also for the other examples presented in this work. \ra{However}, as in this case the network needs to map a more complicated function to distinguish the two polarization modes, it is worth expecting that lager network could perform better, possibly approaching the MCMC accuracy. 

Another important test that can be done with tensor fields projected on the sphere, is the evaluation of parameters from partially covered maps. Ideally, one would expect the uncertainty on cosmological parameter to scale as $f_{sky}^{-0.5}$, at the first order. However, the computation of power spectra on patches of the sky is problematic, as masking induces correlation between Fourier modes and mixing between polarization states. These effects can lead to biases in the estimation of cosmological parameters from spectra, as well as to an increase variance, especially when large scales (low multipoles) are considered. Sophisticated power spectrum estimators, correcting for multipole and polarization state mixing, are currently used to mitigate this problem \citep{Tristram05, Grain09}. It is therefore interesting to understand how the NN accuracy scales as a function of the considered sky fraction.\par

\begin{figure}[!h]
	\centering
	\includegraphics[width=8cm]{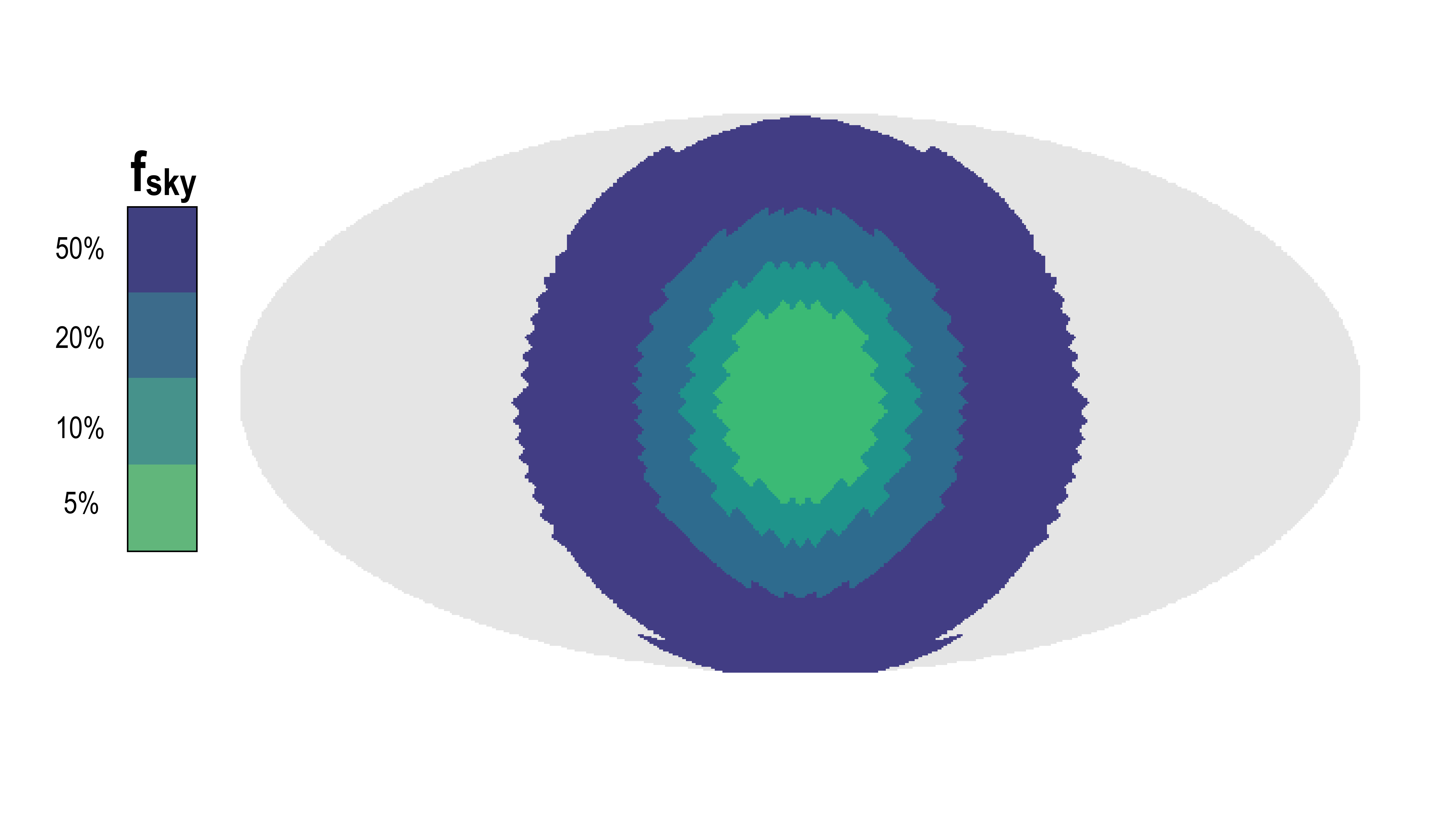}
	\caption{\label{fig:sky_masks} The four circular masks applied to the estimation of the $\ell_p^{E/B}$ parameters (see Section~\ref{sec:tensor_field}), with a retained sky fraction of about 50, 20, 10 and 5\,\%.  On the image masks shown with dark colors retain also the portion of the sky included in those with lighter colors.}
\end{figure}

In our test, we have considered four sky masks, obtained as circular patches in the sky, with retained sky fraction\footnote{Masks are obtained by retaining the portions of the sky included in circular regions, centered on the central pixel of a map at $N_\text{side}=16$, with radii equal to 90, 53, 37 and $26^{\circ}$ and have sky fraction of about 50, 20, 10 and 5\,\% respectively.} varying between $\sim50\,\%$ and $\sim5\,\%$.(Fig.~\ref{fig:sky_masks}). For each of the four masks, we have trained a new CNN, where the input simulated maps have been obtained as described before, but the $Q$/$U$ signal outside the mask have been set to zero. The output of the networks is, as before, the pair of estimates for $\ell_p^E$ and $\ell_p^B$.

Results of are summarized in Table~\ref{tab:gauss_mask}. The mean percentage error on the recovered $\ell_p^{E/B}$ parameter scales roughly as $f_{sky}^{-0.36}$, meaning, therefore, that the NN accuracy is closer to the optimal one for smaller masks (this could be due to the smaller amount of data that it needs to process) and that mixing between different modes and polarization states does not affect its performance. 

\begin{table}[!h]
      \caption{Accuracy (mean percentage error) reached on the estimation of the $\ell_p^{E/B}$ parameters with CNN for the full sky case and the four sky masks shown in Fig.~\ref{fig:sky_masks}.}
         \label{tab:gauss_mask}
     $$ 
         \begin{tabular}{ccccc} 
            \hline\hline
            \noalign{\smallskip}\noalign{\smallskip}
          $f_{sky}=1$ &   $f_{sky} = 0.5$ &  $f_{sky} = 0.2$ &  $f_{sky} = 0.1$ & $f_{sky} = 0.05$\\        
            \noalign{\smallskip}
             \hline
            \noalign{\smallskip}
             \noalign{\smallskip}
               2.7\,\% &3.9\,\% &5.3\,\% & 6.4\,\% & 8.4\,\%       \\                           
            \noalign{\smallskip}
            \hline
         \end{tabular}
     $$ 
      \end{table}

\subsection{Estimation of $\tau$}

In Sect.\ref{sec:tensor_field}, we have showed that our network can estimate the value of simple parameters defined in Fourier space using data in pixel space as input. This proves that the spherical CNN algorithm presented in this paper works well for this kind of regression problems, making it suitable for estimating real cosmological parameters directly from CMB maps (either in total intensity or in polarization). Consequently, we have tested our network architecture on a more realistic case: the estimation of the value of the Thomson scattering optical depth at reionization.

CMB photons, released at the last scattering surface in the early universe, interact with free electrons of the intergalactic medium, ionized by the first emitting objects, at a redshift $z$ between about 11 and 6, during the so-called \emph{reionization epoch}. This causes modifications in the CMB signal both in total intensity and in polarization. In particular, for what concern the polarized signal, the effect is visible especially at the larger angular scales (for multipoles $\ell\lesssim20$) as a bump in the $E$-mode power spectrum, caused by the new Thomson scattering events experienced by the CMB photons. The optical depth of Thomson scattering at reionization, usually indicated with the letter $\tau$, parametrizes the amplitude and shape of this low-ell bump.

In order to constrain the value of $\tau$, high sensitivity polarization observations on large portion of the sky are needed. Moreover, at the large angular scales affected by the reionization bump, spurious signals, coming either from instrumental systematic effects or from residual foreground emission, can strongly contaminate the measurements, making the estimation of this parameter particularly tricky.  Currently, the tightest constraint available on the value of $\tau$ comes from the full sky observation of the \textit{Planck} satellite, with $\tau= 0.054\pm0.007$ (68\,\% confidence level) \citep{PlanckVI18}.  \ra{Among the six cosmological parameters of the standard $\Lambda$CDM model, $\tau$ is the parameter whose value is currently constrained with the largest uncertainty.}

The peculiarity of $\tau$ makes it an interesting test case for parameter estimation using the algorithm presented in this work. Firstly, the fact that $\tau$ affects the large angular scales requires algorithms defined on the sphere,  as the flat-sky approximation cannot be applied. Secondly, since mainly multipoles at $\ell\lesssim20$ are affected by the parameter, low resolution maps (at $N_\text{side}=16$) are sufficient to fit for $\tau$: this makes the problem computationally feasible for CNNs. Lastly, the complexity in the estimation of $\tau$ calls for new analysis techniques that can complement standard parameter estimation routines, thus increasing the confidence on the results.

\begin{figure}[!h]
	\centering
	\includegraphics[width=\columnwidth]{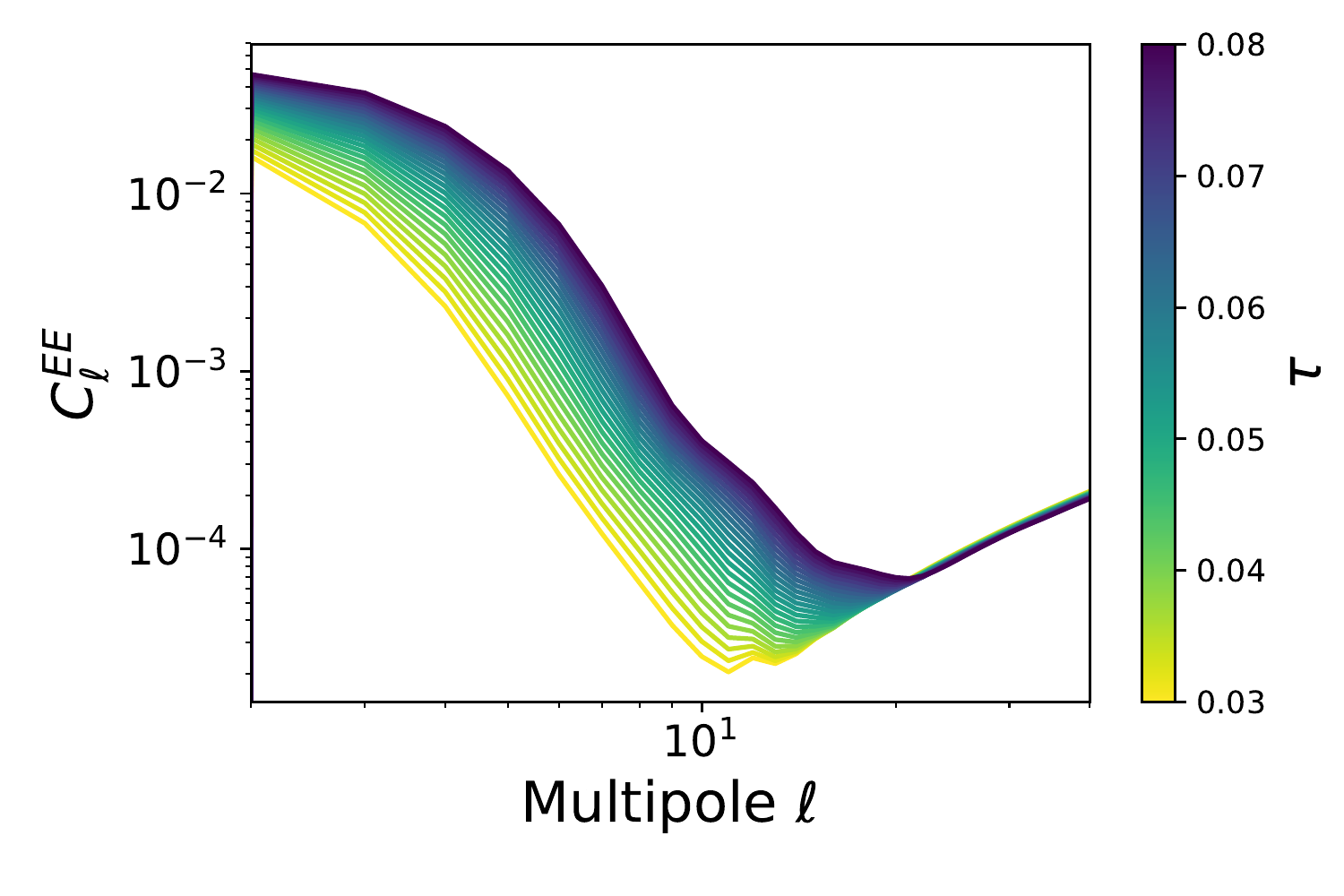}
	\caption{\label{fig:tau_spectra} Subset of the $E$-modes power spectra used to generate to polarization maps for training, validation and test set on which we run the neural network. The value of $\tau$ changes in the range $0.03-0.08$ while all the other cosmological parameters are fixed to the best $\Lambda CDM$ model from Planck results.}
      \end{figure}

To estimate $\tau$ we have used a similar approach as the one described for the $\ell_p$ parameter, generating training, validation, and test sets from simulations. We have computed a set of five thousands $E$-mode power spectra, using the \emph{CAMB} code \citep{Lewis2002}. Each spectrum has a different value of $\tau$, uniformly distributed in the range $0.03-0.08$, while the other cosmological parameters are fixed to the best $\Lambda CDM$ model from Planck \citep{PlanckVI18} (we show a subset of these spectra in Fig.~\ref{fig:tau_spectra}). From these spectra, we have generated 100,000 pairs of full sky $Q$ and $U$ maps with $N_\text{side}=16$ for the training set, and 10,000 and 1,000 for the validation and test sets respectively\footnote{Each pair of maps is simulated by randomly choosing one of the 5,000 spectra with different $\tau$ and with a different seed.}. We have not included any primordial or lensing $B$-mode signal.

We have used the same network architecture, working environment and training procedure described in Sect.~\ref{sec:env}, feeding the network with the simulated $Q$ and $U$ maps and producing an estimate for $\tau$ as the output. By running the CNN on full sky maps, we  have obtained a network accuracy of about 4\,\% (mean percentage error of $\tau$ values over the test set).

We have compared this number with the accuracy reachable through standard MCMC method. As for the previous example, we have maximized the simple chi-squared likelihood of Eq.~\ref{eq:loglike} to fit for the best value of  $\tau$, where the model $\widehat{C}_{\ell}$ represents the theoretical $E$-mode power spectrum, and $\widetilde{C}_{\ell}$ are computed from the test set polarization maps.  The accuracy reached with this method is about 2.8\,\%, a factor $\sim1.5$ better than the NN one.

Although the NN average error is larger, we believe that this result demonstrates that our CNN algorithm could be a feasible approach to constraint the value of $\tau$. Obviously, a complete analysis to understand its real potentiality is needed, including the addition of realistic noise, sky masking, optimization of the NN architecture, error estimation, and so on. However, this is outside the objective of the present work and is hence deferred to a dedicated paper.

\label{sec:tau}

%% file: conclusions.tex
\section{Discussion and conclusions}

\label{sec:conclusions}
In this work, we have presented a novel algorithm for the application of CNNs to signals projected on the sphere. We considered the HEALPix tessellation, which allowed us to easily implement convolutional and pooling layers on a pixel domain. The HEALPix pixelization scheme is used in several scientific fields; therefore, the implementation of efficient and compatible CNN algorithms can be of great interest in several contexts.

Our algorithm presents some main advantages. Firstly, \ra{convolution is done in the pixel domain, as for standard 1D Euclidean cases:} this allows the construction, training, and testing of the CNNs using the existing highly-optimized neural network libraries, such as PyTorch and TensorFlow. Moreover, \ra{the algorithm is computationally efficient, with a computational complexity that scales as $O(N_{pix})$.} Lastly, \ra{the process can be easily optimized to run only on portion of the sphere.}

\ra{On the other hand and contrary to other spherical CNNs recently introduced in literature  \citep{Perraudin2018, Cohen2018}, our implementation is not rotationally equivariant. Although this makes the network sub-optimal, the good results obtained on our applications show that training makes the CNN able to overcame this deficiency.}

We have built a simple network architecture, which includes four convolutional and pooling layers. Therefore, convolution is applied to maps at decreasingly lower resolutions, making the network sensitive to feature at different angular scales. The same architecture has been \ra{used} for all the applications presented in this work.

We have first tested the feasibility of our approach on the recognition of handwritten digits (MNIST dataset) projected on the HEALPix sphere. Results show that the trained CNN can recognize handwritten digits that cover large portion of the sphere, with an accuracy of about 95\,\%, a performance comparable with other kinds of spherical NNs. As our network is not spherical invariant, it must be trained with images projected at different positions and orientations on the sphere. When properly done, the performance of the CNN is independent on the position and \ra{orientation} of images.

We \ra{have then moved} to applications related to the Cosmology field. In particular, we have applied CNNs to the estimation of cosmological parameters from CMB simulated observations, \ra{impacting the very large angular scales ($\ell\lesssim20$)}, directly from temperature or polarization sky maps. \ra{This work represents the first one in which NNs are tested on such a task.}

We have started with the estimation of a mock parameter, $\ell_p$, defining the angular scales at which the power spectrum of a Gaussian field projected on the sphere peaks. We have used the same network architecture applied the MNIST recognition, changing only input and output layers depending on whether we fit for the value of  $\ell_p$ in temperature or polarization.

For the temperature case (scalar field projected on the sphere) we have applied the network to both noiseless or noisy maps, showing that it is able to correctly retrieve the value of $\ell_p$. The  reached accuracy is comparable with the one obtained from the estimation with standard bayesian methods applied to the angular power spectra computed from maps. \ra{In particular, for the noiseless case, CNNs preform about a factor two worse then standard methods, but this factor move close to one for the noisy maps.}

We have applied the same CNN architecture also to maps in polarization (tensor field projected on the sphere). In this case we fed the network with pairs of Stokes $Q$ and $U$ maps, retrieving the values of two parameters $\ell_p^E$ and $\ell_p^B$, representing the angular scales at which the polarization $E$ and $B$-mode power spectra peak. The accuracy reached is about four times worse than for the standard bayesian estimation (percentage error on the retrieved $\ell_p^{E/B}$ values is $\sim2.7\%$ against $\sim0.7\%$).  \ra{Although not being optimal,} this result is indeed remarkable, as it proves for the first time that CNNs are able to distinguish and separate E/B polarization in pixel space. We have repeated the test in case of partial sky coverage, applying sky masks, with retained sky fraction between 50 and 5\%, to the input $Q$ and $U$ maps. The network performs well also under these conditions, with achieved accuracy closer to the optimal one for the smaller masks. This shows that mixing between multipoles and polarization states, \ra{due to} the sky cut, does not impact the network performance.

Lastly, we have tested the network on the estimation of a real cosmological parameter: the optical depth of Thomson scattering at reionization ($\tau$). \ra{ Although in this work we only present preliminary results on the estimation of $\tau$ as a proof-of-concept of our approach, this represents one possible concrete and important application of our network, to be fully explored in the future. The $\tau$} parameter mostly impacts the large angular scales of $E$-mode polarized emission. Its estimation is complicated by the fact that spurious signals, coming from instrumental systematic effects or Galactic residual emissions, can strongly contaminate the cosmological one at low multipoles. As a matter of fact, among the six cosmological parameters of the standard $\Lambda$CDM model, $\tau$ is the parameter whose value is currently constrained with the largest uncertainty \citep{PlanckVI18}. For these reasons, pairing standard estimation methods with new ones based on NNs, would be of great importance to achieve a better constraint and cross check results.

We have applied the same CNN architecture used in the previous examples to simulated maps, with values of $\tau$ in the range $0.03-0.08$. We have tested the performance of the network in the simplest case of full sky and noiseless observations, and without applying any specific optimization to the CNN architecture. Results shows that the network is able to reach an accuracy on the estimation of $\tau$ that is $\sim1.5$ worse than standard Bayesian fitting.

Our findings represent a first step towards the possibility to estimate $\tau$ with NNs, and show that our implementation of CNNs on the sphere \ra{is} a valid tool to achieve the goal.  Obviously, a more sophisticated analysis is needed, including complications to make simulations more representative of real data, such as the introduction of realistic noise, instrumental systematic effects and foreground signals. In order to have a reliable estimate of $\tau$, a full characterization if the results obtained with NNs is also needed, especially regarding the estimation of uncertainties. Moreover, specific optimization of the CNN architecture is necessary. We defer all these studies to a subsequent work. 
 \par

%% file: appendix_striped_maps.tex
\section{Detecting stripes in a map}
\label{sec:exampleStripes}

\begin{figure}[!h]
	\centering
	\includegraphics[width=0.6\columnwidth]{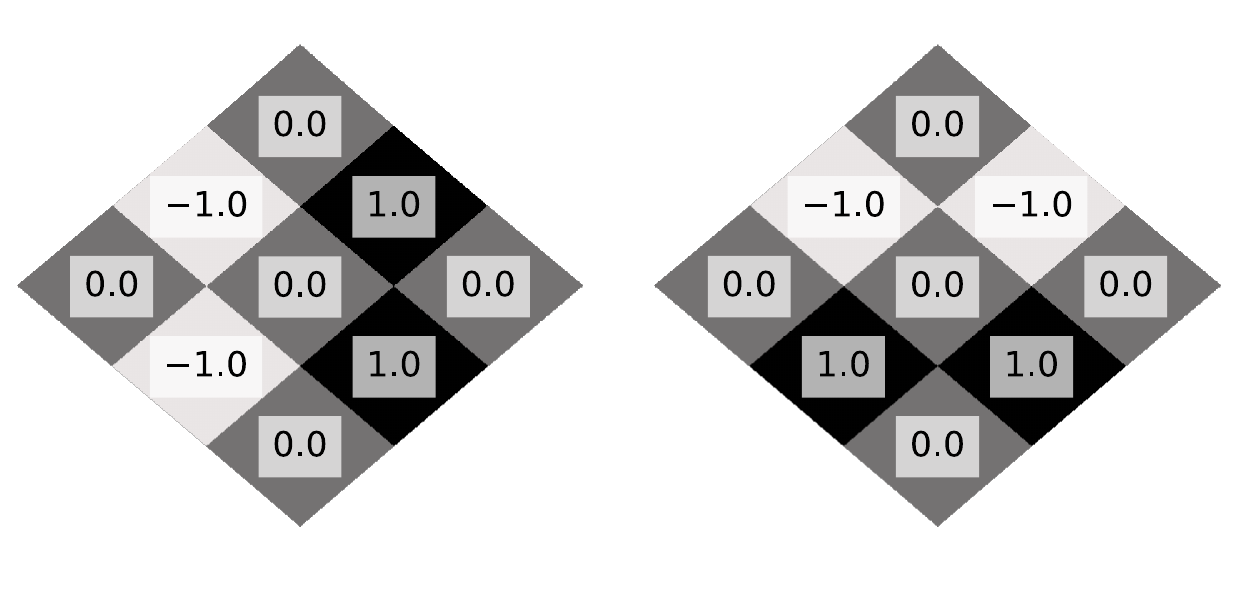}
	\caption{\label{fig:stripedFilters}Filters used in the example of convolution described in Sect.~\protect\ref{sec:convolutionSphere}. Black pixels are set to 1, white pixels to $-1$, gray pixels to zero. The filter on the left is designed to pick vertical features, i.e., stripes that are aligned with meridians. The filter on the right picks features running along parallels. These filters are here shown with a diamond shape, but practically they are applied as 1D vectors to maps, following the algorithm sketched in Fig.~\protect\ref{fig:filterOutline}. The unrolling into vectors is done clockwise, starting from the NW direction.}
\end{figure}

\begin{figure}[!h]
	\centering
	\includegraphics[width=\columnwidth]{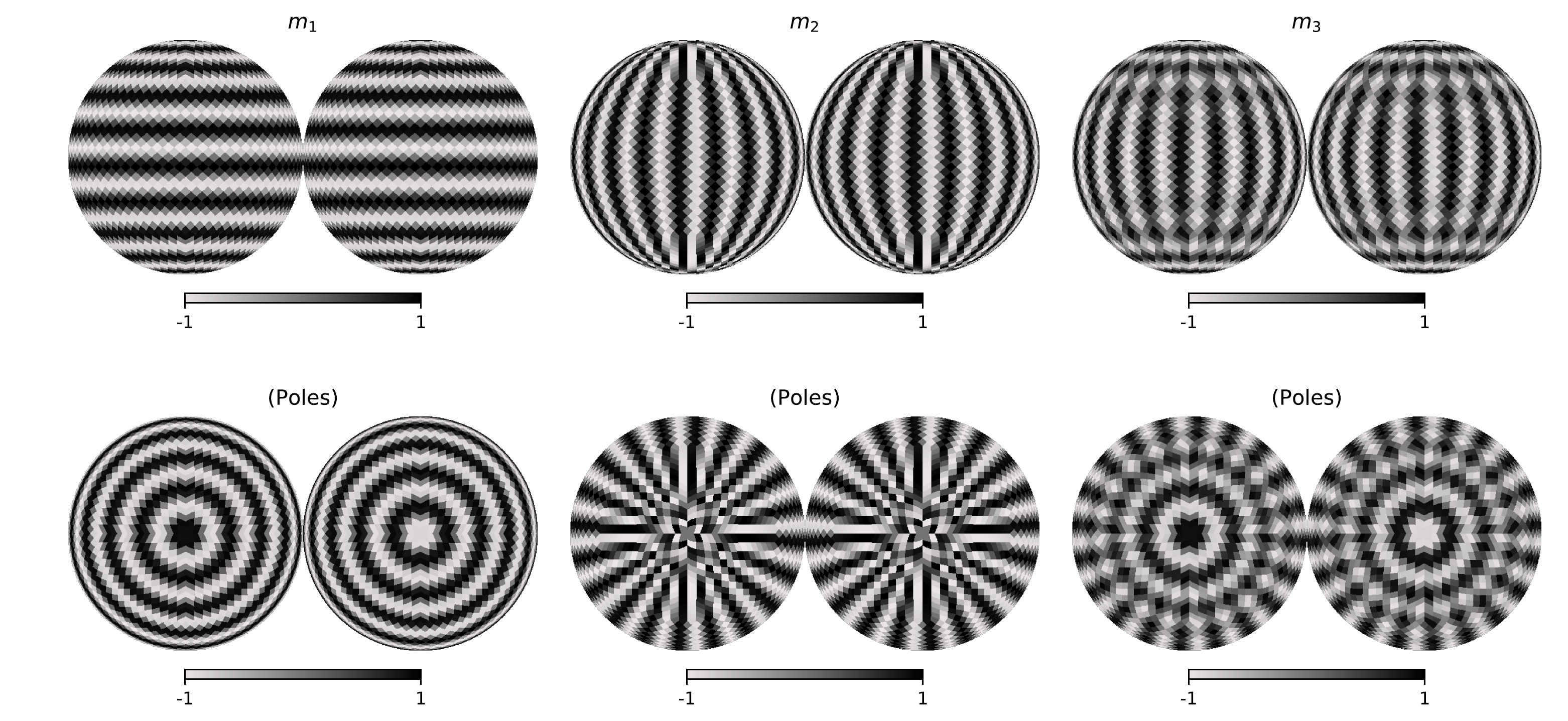}
	\caption{\label{fig:stripedMapsOverview} Maps used in the explanation of convolution in Sec.~\protect\ref{sec:convolutionSphere}. All the maps have $N_\text{side}=16$. Each map is shown using two orthographic projections: in the upper row, maps are represented as seen from the Equatorial plane; in the lower row, maps are centered around the poles.}
\end{figure}

\begin{figure}[!h]
	\centering
	\includegraphics[width=\columnwidth]{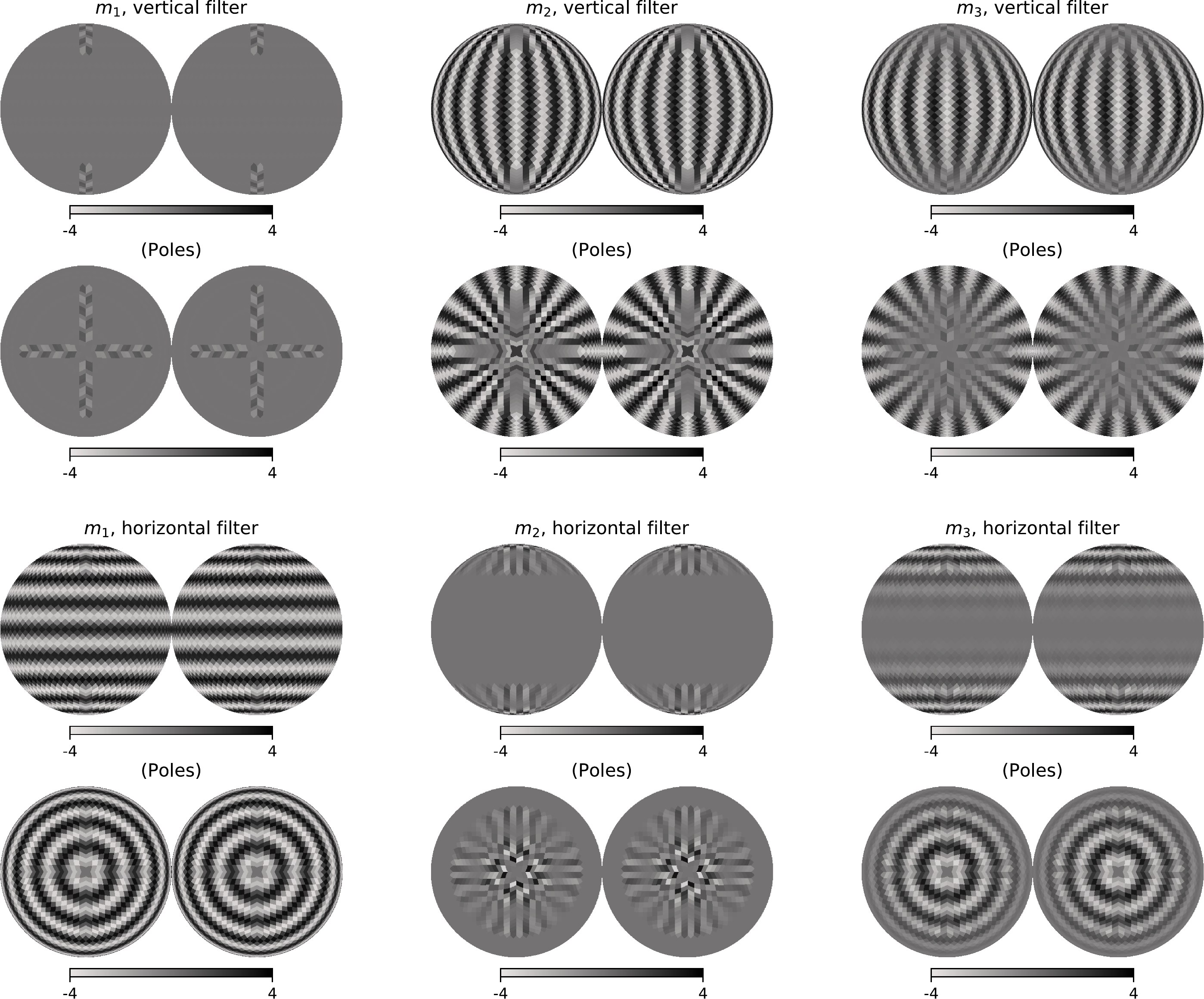}
	\caption{\label{fig:stripedMapsFiltered} Result of the application of the two filters in Fig.~\ref{fig:stripedFilters} to the three maps shown in Fig.~\protect\ref{fig:stripedMapsOverview}.}
\end{figure}

\ra{In this appendix, we show a pedagogical application of the convolution algorithm presented in Sect.~\ref{sec:convolutionSphere}.} We have created two nine-elements filters, shown in Fig.~\ref{fig:stripedFilters}. In this figure, filters are shown with a diamond shape, but they are flattened into a nine-element vector to compute the convolution, following the clockwise order of elements. The sum of the values of the weights $w_i$ is zero in both filters, and the only non-zero pixels are aligned along vertical or horizontal lines.  In this way, the application of a filter to a block of nine pixels on a map will be significantly different from zero only if the pixels in this block show some vertical or horizontal features.

We have produced three maps $m_1$, $m_2$, $m_3$, where the value associated with each pixel $p$ is given, respectively, by: 
\begin{align}
	\label{eq:stripedMapI} p_1(\theta, \varphi) &= \sin(20\,\theta),\\
	\label{eq:stripedMapII} p_2(\theta, \varphi) &= \sin(20\,\varphi),\\
	\label{eq:stripedMapIII} p_3(\theta, \varphi) &= p_1(\theta, \varphi)\,\sin^2(\theta) + p_2(\theta, \varphi)\,\cos^2(\theta),
\end{align}
with $(\theta, \varphi)$ being the colatitude and longitude of the pixel center ($0 \leq \theta \leq \pi$, $0 \leq \varphi < 2\pi$). The maps are shown in Fig.~\ref{fig:stripedMapsOverview}; map $m_3$ is a weighted combination of $m_1$ and $m_2$, where the weights $\sin^2(\theta)$ and $\cos^2(\theta)$ make horizontal stripes (from $m_1$) and vertical stripes (from $m_2$) negligible at the Equator and at the poles, respectively.

The application of the horizontal and vertical filters in Fig.~\ref{fig:stripedFilters} to the maps in Fig.~\ref{fig:stripedMapsOverview} is shown in Fig.~\ref{fig:stripedMapsFiltered}. The result of applying the horizontal filter to the vertically striped map is a map with almost no feature: the only residuals are shaped like a plus sign around the pole and correspond to those pixels that only have 7 neighbors instead of 8 (see Fig.~\ref{fig:pixelNeighbors}), as the sum of the weights is no longer zero in this case. Similar results occur in the other maps. In the the case of map $m_3$, which contains both horizontal and vertical stripes, the two filters correctly pick the right feature depending on the latitude, as expected.